%% file: main_arxiv.tex
\PassOptionsToPackage{dvipsnames}{xcolor}

\documentclass[11pt,letterpaper]{style}

\usepackage[numbers]{natbib}
\usepackage{graphicx}
\usepackage{booktabs}
\usepackage{amsmath,amsfonts,amssymb}
\usepackage{cleveref}
\usepackage{subcaption}
\usepackage{wrapfig}
\usepackage{multirow}
\usepackage{colortbl}
\usepackage{listings}
\usepackage{xparse}
\usepackage{fontawesome5}
\usepackage{bxcoloremoji}
\usepackage{float}
\usepackage{placeins}
\usepackage{threeparttable}

\graphicspath{{./}{fig/}{figures/}{plot/}{pdf/}{table/}}
\usepackage{amsthm}
\usepackage{tcolorbox}
\usepackage{svg}
\tcbuselibrary{skins,breakable}
\tcbuselibrary{listingsutf8}
\usepackage{titletoc}

\usepackage{setspace}
\usepackage{pifont}
\usepackage{mathtools}
\usepackage{enumitem}
\usepackage{arydshln}
\usepackage{bbm}
\usepackage{lineno}
\usepackage{makecell}
\usepackage{adjustbox}
\usepackage{algorithm}
\usepackage{algorithmic}
\usepackage{caption} 
\usepackage{wrapfig}
\usepackage{hyperref}
\usepackage{bbding}

\hypersetup{colorlinks=true,linkcolor=red,urlcolor=blue,citecolor={blue}}

\definecolor{mygray}{gray}{0.9}
\definecolor{syncol}{RGB}{243,246,249}
\definecolor{wildcol}{RGB}{215,240,235}
\definecolor{drop1}{RGB}{180,225,220}
\definecolor{drop2}{RGB}{150,210,200}
\definecolor{drop3}{RGB}{120,195,185}
\definecolor{drop4}{RGB}{95,180,170}
\definecolor{drop5}{RGB}{65,160,150}
\definecolor{lightblue}{RGB}{26,82,249}

\definecolor{lightgray}{gray}{0.9}

\definecolor{myblue1}{HTML}{0171DC}
\definecolor{myblue2}{HTML}{013978}

\NewDocumentEnvironment{minted}{O{} m +b}{%
}{}

\newcommand{\afficon}[1]{\textsuperscript{\fontsize{9pt}{9pt}\selectfont #1}}
\newcommand{\equal}{\textsuperscript{*}}     
\newcommand{\advisor}{\textsuperscript{\dag}}

\renewcommand\Authfont{\centering\normalfont\bfseries\fontsize{11}{15}\selectfont}
\renewcommand\Affilfont{\centering\normalfont\fontsize{10}{15}\selectfont}

\title{FUAS-Agents: Autonomous Multi-Modal LLM Agents for Treatment Planning in Focused Ultrasound Ablation Surgery}
\runningtitle{FUAS-Agents: Autonomous Multi-Modal LLM Agents for Treatment Planning in Focused Ultrasound Ablation Surgery}

\author{%
    {\Authfont
    \textbf{Lina Zhao}\equal\afficon{1} \quad
    \textbf{Zihao Bian}\equal\afficon{1} \quad
    \textbf{Qingyue Chen}\equal\afficon{2} \quad
    \textbf{Yafang Li}\afficon{1} \\
    \textbf{Zhiyi Luo}\afficon{1} \quad
    \textbf{Jiaxing Bai}\afficon{3} \quad
    \textbf{Guangbo Li}\afficon{4} \quad
    \textbf{Min He}\afficon{5} \\
    \textbf{Kezhi Li}\advisor\afficon{6} \quad
    \textbf{Huaiyuan Yao}\advisor\afficon{3} \quad
    \textbf{Zongjiu Zhang}\advisor\afficon{1}
    }\\
    \vspace{0.2cm}
    {\Affilfont
    \textsuperscript{1} Tsinghua University \quad
    \textsuperscript{2} Wuhan University \quad
    \textsuperscript{3} Xi'an Jiaotong University \\
    \textsuperscript{4} Beijing Jiaotong University \quad
    \textsuperscript{5} Chongqing Haifu Hospital \\
    \textsuperscript{6} University College London, UCL Institute of Health Informatics \\
    \vspace{0.1cm}
    \texttt{\{zhaoln23, bianzh24, luozy24\}@mails.tsinghua.edu.cn, zhangzongjiu@tsinghua.edu.cn} \\
    \texttt{alyssa@whu.edu.cn, liyafangnku@163.com, yiyeyikezhiqiu@gmail.com} \\
    \texttt{20120009@bjtu.edu.cn, hemin0x1@163.com, ken.li@ucl.ac.uk, huaiyuanyao@gmail.com} \\
    \vspace{0.1cm}
    \equal \ Equal Contribution \quad \advisor Corresponding Author
    }
}

\begin{document}

\input{sections/0abstract}

\newcommand{\TitleLinks}{%
\centering
    \vspace{8pt}
}
\maketitle

\input{sections/1intro}

\input{sections/2result}

\input{sections/3discussion}
\input{sections/4method}

\bibliographystyle{unsrtnat}  
\bibliography{references}

\appendix

\input{sections/5appendix}

\end{document}

%% file: sections/0abstract.tex
\begin{abstract}
Focused Ultrasound Ablation Surgery (FUAS) has emerged as a promising non-invasive therapeutic modality, valued for its safety and precision. Nevertheless, its clinical implementation entails intricate tasks such as multimodal image interpretation, personalized dose planning, and real-time intraoperative decision-making processes that demand intelligent assistance to improve efficiency and reliability.
We introduce FUAS-Agents, an autonomous agent system that leverages the multimodal understanding and tool-using capabilities of large language models (LLMs). The system was developed using a large-scale, multicenter, multimodal clinical dataset of over 3000 cases from three medical institutions. By integrating patient profiles and MRI data, FUAS-Agents orchestrates a suite of specialized medical AI tools, including segmentation, treatment dose prediction, and clinical guideline retrieval, to generate personalized treatment plans comprising MRI image, dose parameters, and therapeutic strategies. The system also incorporates an internal quality control and reflection mechanism, ensuring consistency and robustness of the outputs.
We evaluate the system in a uterine fibroid treatment scenario. Human assessment by four senior FUAS experts indicates that 82.5\%, 82.5\%, 87.5\%, and 97.5\% of the generated plans were rated 4 or above (on a 5-point scale) in terms of completeness, accuracy, fluency, and clinical compliance, respectively. In addition, we have conducted ablation studies to systematically examine the contribution of each component to the overall performance. These results demonstrate the potential of LLM-driven agents in enhancing decision-making across complex clinical workflows, and exemplify a translational paradigm that combines general-purpose models with specialized expert systems to solve practical challenges in vertical healthcare domains.

\end{abstract}


%% file: sections/1intro.tex
\section{Introduction}

Focused ultrasound ablation surgery (FUAS) is a non-invasive technique that employs focused ultrasound to induce coagulative necrosis in targeted tissues through thermal and cavitation effects. Clinical evidence supports its efficacy in treating a wide range of benign and malignant solid tumors, such as uterine fibroids, hepatocellular carcinoma, pancreatic and prostate cancers, and breast fibromas, as well as certain non-neoplastic diseases \cite{ChineseMedicalAssociation2020}. Compared to conventional surgery, FUAS offers advantages including reduced trauma, faster recovery, and fewer complications\cite{2021High}, making it increasingly attractive to clinicians and patients. Nevertheless, challenges remain, including limitations in multimodal image guidance, reliance on operator experience for dose determination, and the lack of individualized treatment planning\cite{Zhou2017Noninvasive}, which hinder broader clinical adoption and application.


Previous studies have investigated the integration of AI techniques into FUAS to address the challenges faced in the aforementioned clinical processing. In image processing,  ~\cite{2018Deformable} combined GCN with the DMAC model to perform automatic preoperative lesion segmentation. ~\cite{2020HIFUNet} proposed a 3D convolutional neural network-based framework for non-rigid registration between MRI and ultrasound images using unsupervised learning. Intraoperative monitoring based on deep learning techniques is separately explored by ~\cite{2023Diffusion} and ~\cite{2020Real} to support clinical decision-making. In terms of dose prediction, models such as MLP and XGBoost have been developed to estimate therapeutic parameters~\cite{hu2023machine}, while others employed Deep Multimodal Teacher-Student (MMTS) frameworks to reconstruct temperature distributions from ultrasound echo signals, enabling thermal-feedback-driven dose prediction~\cite{luan2024real}. In terms of treatment efficacy, machine learning models have also been applied to predict non-perfusion volume reduction and residual tissue regeneration following FUAS~\cite{zhang2022magnetic}.


Although existing AI-based approaches have made initial attempts to address key challenges in FUAS, which are largely based on task-specific, expert-driven models. These models often suffer from limited generalization, heavy dependence on annotated data, and poor adaptability to complex clinical settings, thereby limiting the intelligent development and broader adoption of FUAS technology. This underscores the urgent need for a more powerful and unified framework capable of multimodal semantic understanding, autonomous reasoning, and cross-task generalization to advance automated and personalized treatment planning~\cite{zhang2025revolutionizing}~\cite{alsaad2024multimodal}.


In recent years, multimodal large language models (MM-LLMs) and agent technologies driven by them have rapidly evolved~\cite{kaczmarczyk2024evaluating, chen2026hipo, chen2026fastsafeonlinereinforcement} and have been increasingly introduced into a wide range of medical applications~\cite{thirunavukarasu2023llmmedicine}, including clinical diagnosis~\cite{2024MMedAgent}~\cite{2024ClinicalLab}, decision support~\cite{tang2023medagents}~\cite{2024MDAgents}, medical report generation~\cite{jin2026grounded}~\cite{chen2401chexagent}~\cite{sharma2024cxr}, medical education~\cite{2024AIPatient}~\cite{2024Benchmarking}, and healthcare management~\cite{2024Depression}~\cite{2024Polaris}. Owing to their capabilities in natural language understanding~\cite{Zhao_Zhao_Song_He_Zhang_Zhang_Li_2026, chen2026conformalfeedbackalignmentquantifying}, multimodal semantic alignment~\cite{zhao2025secondorderfinetuningpainllmsa, chen-etal-2025-unveiling-privacy}, knowledge representation~\cite{liu2025mcqaevalefficientconfidenceevaluation}, autonomous reasoning~\cite{lilodriver}, and tool utilization~\cite{instructional}~\cite{2024A}~\cite{yao2025comalcollaborativemultiagentlarge}, LLM-based medical agents are widely regarded as a promising technological pathway for addressing complex clinical problems~\cite{2025A}~\cite{yang2023large}~\cite{single_cell}.

Recent research has begun to explore treatment plan generation. For example, early radiotherapy planning agents based on cGAN attempted to achieve automated planning~\cite{0An}, but their capabilities in complex language understanding and multimodal information fusion were limited due to a lack of underlying model support. Subsequently, some works further introduced underlying models, retrieval-enhanced generation (RAG), and reinforcement learning mechanisms to improve radiotherapy planning by simulating multi-role collaboration or closed-loop optimization~\cite{nusrat2025autonomous}. Other studies have integrated agents with medical tools for personalized treatment recommendations~\cite{2025TxAgent}, or adopted a closed-loop decision framework to formulate tumor treatment plans through observation, simulation, and policy optimization~\cite{raiaan2024review}.

However, the transferability of these approaches to focused ultrasound ablation—a precision treatment scenario that relies heavily on multimodal imaging, fine-grained dose control, and individualized anatomical structures—remains limited. Consequently, how to construct a multimodal agent system tailored for FUAS treatment that can operate under real clinical constraints, alleviate practical clinical challenges, and generate interpretable treatment plans has become the central focus of this study. To this end, we propose FUAS-Agents, a multi-agent system designed for focused ultrasound ablation treatment planning. Within a unified framework, the system integrates medical image analysis, radiomics modeling, and machine learning techniques to enable MRI image segmentation, treatment dose prediction, and personalized treatment strategy generation. FUAS-Agents is developed using real-world clinical data from over 3,000 cases across multiple centers, and adopts an MM-LLM–based multi-agent architecture in which different agents are responsible for task planning, tool invocation, strategy generation, memory retrieval, and result optimization, thereby simulating the multi-step decision-making process observed in real clinical practice.

The main contributions of this study can be summarized as follows: (1) From the perspective of clinical decision-making in FUAS practice, we develop an end-to-end multi-agent decision-support system that assists clinicians in treatment planning by systematically addressing key real-world challenges, including strong dependence on individual operator experience, limited personalization across patients, and difficulties in achieving consistent and standardized treatment strategies. By generating patient-specific and interpretable treatment plans, the system provides practical support for clinicians in making more informed and reproducible decisions. (2) To meet the safety and reliability requirements of clinical use, we propose a clinically oriented collaborative architecture that integrates large foundation models with domain-specific expert models. This design enhances the system’s reasoning and decision-making capabilities while ensuring that generated recommendations remain aligned with established medical knowledge and clinical constraints, thereby increasing clinicians’ trust and confidence in the system outputs. (3) Beyond the specific FUAS application, We provide a reusable design paradigm for vertical medical agent systems, demonstrating that reliable clinical-grade systems require more than general-purpose LLMs and prompt engineering; effective solutions must integrate foundation models with domain knowledge, specialized tools, and task-specific fine-tuning.

To assess the effectiveness and clinical relevance of FUAS-Agents, we conducted a series of evaluations in the context of uterine fibroid treatment. Comparative experiments against representative large-scale medical language models, as well as ablation studies, were both evaluated by four senior FUAS specialists. The experts assessed the generated treatment plans across four dimensions: completeness, accuracy, fluency, and clinical compliance. The results indicate that FUAS-Agents achieved high performance, with 82.5\%, 82.5\%, 87.5\%, and 97.5\% of the plans receiving scores of 4 or higher on a 5-point Likert scale, demonstrating the system’s robustness and potential for clinical applicability.

%% file: sections/2result.tex


\section{Result}

\subsection{Overview of the FUAS-Agents Framework}
The FUAS-Agents framework is a modular multi-agent system engineered to automate personalized treatment planning for FUAS. Inspired by clinical cognitive processes, the system simulates the physician’s reasoning workflow—encompassing image interpretation, dose estimation, and strategy formulation—through collaborative, knowledge-based interaction. The architecture adheres to an "Agent–Tool–Memory" paradigm comprising three core components: (1) Agents, responsible for high-level reasoning, task planning, and validation; (2) Tool Modules, which provide deterministic capabilities such as precise image segmentation and dose prediction; and (3) a Memory Module, a shared knowledge base containing clinical guidelines and prior cases, accessed by all agents via a RAG interface.

\begin{figure}[h]
\centering
\includegraphics[width=0.9\textwidth]{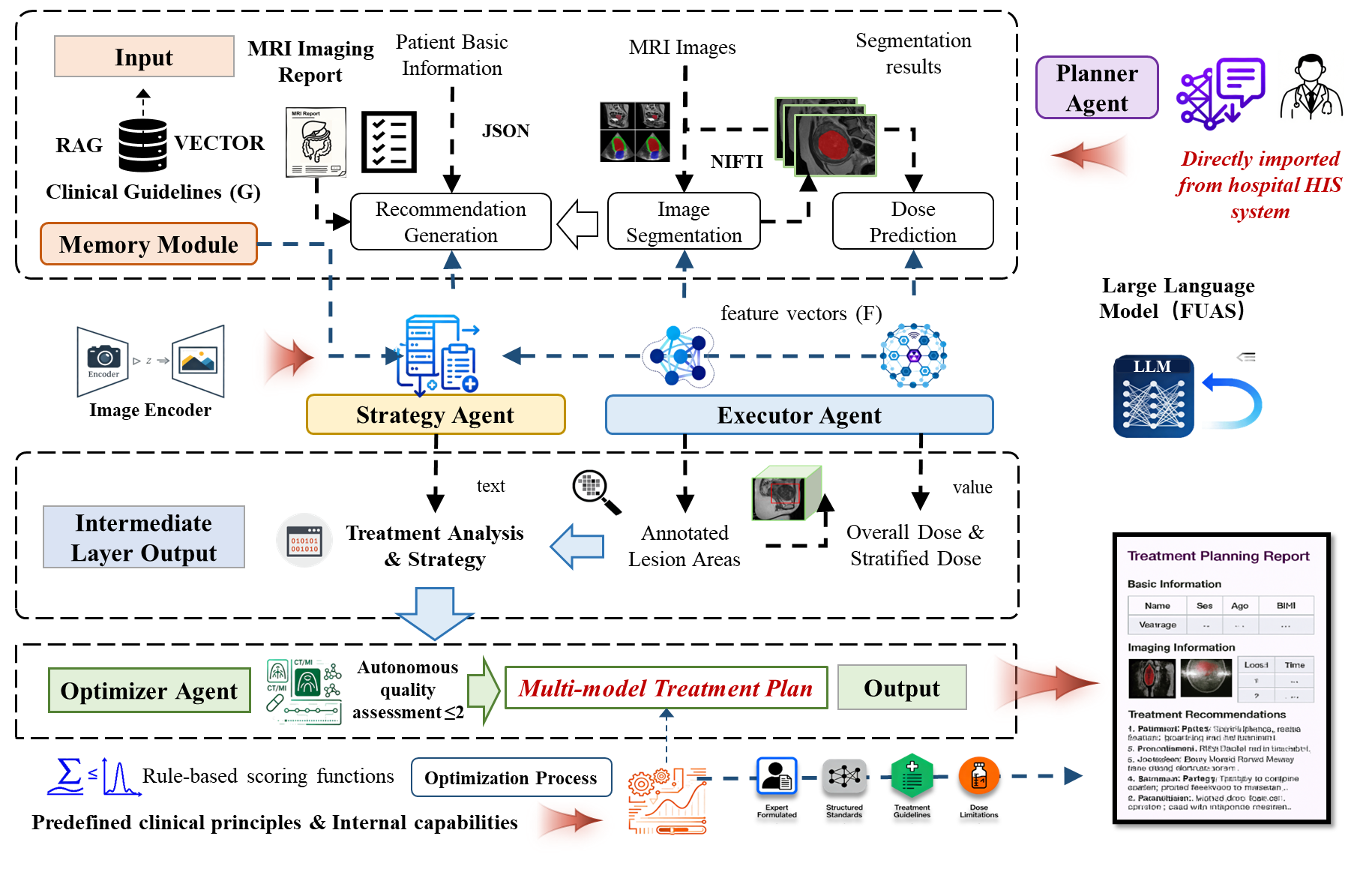}
\captionsetup{justification=justified, singlelinecheck=false}
\caption{Overview of FUASAgents’ data flow and data format. The framework illustrates the end-to-end data flow and agent collaboration for FUAS treatment planning. Patient-specific inputs, including MRI imaging reports, basic clinical information, and raw MRI images imported from the hospital HIS system, are first processed by the Planner Agent, which coordinates task decomposition and workflow scheduling. MRI images are encoded and passed to the Executor Agent for image segmentation and dose prediction, producing lesion annotations, feature vectors, and stratified dose estimates. Based on these intermediate outputs, the Strategy Agent synthesizes treatment analysis and personalized planning strategies in textual form. An Optimizer Agent subsequently performs autonomous quality assessment and optimization using rule-based clinical constraints and predefined principles. The final output is a structured treatment planning report integrating imaging results, dose recommendations, and treatment strategy.}\label{fig:workflow}
\end{figure}

A detailed workflow focusing on the data flow and data formats is presented in Figure~\ref{fig:workflow}. The process initiates with the Planner Agent, which decomposes complex surgical requirements into executable sub-tasks and coordinates downstream modules. Upon receiving instructions, the Executor Agent selects the appropriate tools; specifically, the Segmentation and Dose Prediction Tools extract critical imaging features and quantitative parameters. These structured data are subsequently passed to the Strategy Agent to generate a preliminary patient-specific treatment plan. Finally, the Optimizer Agent conducts symbolic validation, triggering a reflection mechanism for immediate plan revision if inconsistencies are detected. This collaborative loop ensures that the final strategy is rigorously aligned with the medical evidence stored in the shared knowledge base.

\subsection{Study Cohorts and Data Characteristics}

To validate the robustness and clinical utility of the framework, we constructed a comprehensive, multi-source dataset comprising multimodal MRI data and clinical records, which were stratified into three task-specific cohorts. For the lesion segmentation task, we analyzed 702 patients with 3D MRI scans and expert-verified manual segmentation masks, partitioning them into a training set ($n=561$) and a validation set ($n=141$) with an 8:2 split to ensure comparable difficulty distributions. The dose prediction cohort analyzed T2-weighted imaging (T2WI) sequences and corresponding device-induced dose records from 210 expert-annotated patients undergoing FUAS treatment. Cases were excluded if they had poor image quality, concomitant use of auxiliary agents (such as absolute ethanol), a postoperative non-perfused volume ratio (NPV) of less than 70\%, or other uterine or adnexal pathologies. The mean age of the included patients was 41.54 ± 6.63 years. The dataset was randomly divided into a training set and a validation set at a ratio of 8:2. No significant difference in dose distribution was observed between the two sets (P = 0.64, Kolmogorov–Smirnov test).  Finally, for treatment strategy generation, we curated a large-scale dataset of over 2,000 clinical records from three collaborating medical institutions. These anonymized records, reviewed by five senior FUAS experts, were utilized to fine-tune the LLM components.

\subsection{Performance of Intelligent Lesion Segmentation and Definition}
The segmentation efficacy of the Executor Agent was rigorously evaluated on the validation cohort using MedSAM-2 as the foundational framework. Quantitative analysis demonstrated that our fine-tuned model achieved substantial superiority over the baseline across all interaction modalities (Table~\ref{tab:seg_performance}). In the fully autonomous mode, the system attained a Dice Similarity Coefficient (Dice) of 0.6645, representing a 4.5-fold improvement over the baseline MedSAM-2. This significant gain indicates a strong zero-shot generalization capability, making it highly effective for rapid, automated lesion screening. When human-in-the-loop interaction was introduced via the "Click" mode (1–3 clicks), the precision further ascended to a Dice score of 0.8550 (+11.5\% vs. baseline). This modality offers an optimal trade-off between clinical efficiency and the high-fidelity boundary definition required for complex anatomical environments. Additionally, the "BBox" mode yielded consistent results with a Dice score of 0.7724, confirming that the Executor Agent maintains robust performance across varying user workflows. Detailed visual comparisons and segmentation metrics are provided in Appendix A.

\setlength{\tabcolsep}{20pt}
\begin{table}[htbp]
\centering
\captionsetup{justification=justified, singlelinecheck=false} 
\caption{Quantitative evaluation of MedSAM-2 and our method across different prompt types. 
Our method consistently outperforms MedSAM-2 across all prompt types, 
with the most significant improvement observed for the ``Autonomy'' and ``Click'' prompt types. 
The bolded numbers indicate the best performance for each metric.}
\label{tab:seg_performance}
\begin{tabular}{llccc}
\toprule
\textbf{Prompt Type} & \textbf{Model} & \textbf{Dice} & \textbf{IoU} & \textbf{Average} \\
\midrule
Autonomy & MedSAM-2 & 0.1577 & 0.1170 & 0.1374 \\
Autonomy & Ours     & 0.6645 & 0.6245 & 0.6445 \\
\midrule
Click    & MedSAM-2 & 0.7670 & 0.7245 & 0.7458 \\
Click    & Ours     & \textbf{0.8550} & \textbf{0.8085} & \textbf{0.8318} \\
\midrule
BBox     & MedSAM-2 & 0.7596 & 0.7391 & 0.7494 \\
BBox     & Ours     & 0.7724 & 0.7574 & 0.7649 \\
\bottomrule
\end{tabular}
\end{table}

\subsection{Precision Radiomics for Individualized Dose Prediction}
Within the FUAS-Agents system, the Executor Agent invoked a combined dose prediction model integrating radiomic features and clinical variables to generate individualized treatment dose recommendations. The combined model demonstrated strong discriminative performance in the training cohort, achieving an area under the receiver operating characteristic curve (AUC) of 0.983, and retained moderate generalization ability in the independent testing cohort (AUC = 0.714). Receiver operating characteristic and decision curve analyses indicated that the model provided positive net clinical benefit across multiple clinically relevant threshold probabilities, suggesting its potential value for decision support in FUAS dose planning (Figure 2).
In the combined model, radiomic features contributed substantially to dose prediction. These features included first-order statistical metrics (such as low gray-level percentiles and energy-related indices), which reflect the overall signal intensity and energy distribution within the lesion, as well as texture features derived from Gray-Level Co-occurrence Matrix (GLCM), Gray-Level Run Length Matrix (GLLM), Gray-Level Size Zone Matrix (GLSZM), Gray-Level Dependence Matrix (GLDM), and Neighboring Gray Tone Difference Matrix (NGTDM) matrices, characterizing gray-level distribution patterns and intra-lesional heterogeneity at different spatial scales. In addition, several clinically relevant variables were incorporated into the combined model, including body mass index (BMI), abdominal wall–related parameters, and the preoperative surgical score. These variables represent patient body habitus, acoustic pathway conditions, and clinically assessed treatment difficulty in FUAS, respectively, and provide important complementary information to image-driven dose prediction.

\begin{figure}[htbp]
    \centering
    \includegraphics[width=0.7\linewidth]{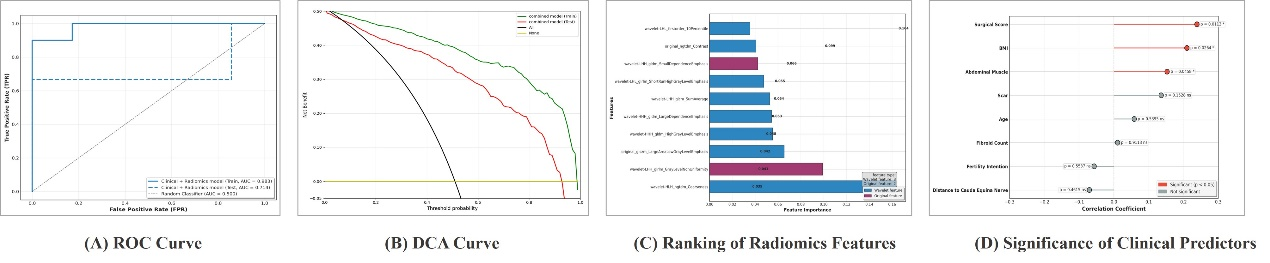}
\captionsetup{justification=justified, singlelinecheck=false}
    \caption{\textbf{Performance evaluation and feature analysis of the dose prediction model.} (A) Receiver operating characteristic (ROC) curve of the combined model, illustrating its discriminative performance; (B) Decision curve analysis (DCA) of the combined model, demonstrating the net clinical benefit across different threshold probabilities; (C) Importance ranking of key radiomics features in the radiomics-based model; (D) Statistical significance of clinical predictors in the clinical prediction model.}
    \label{fig:dose_prediction}
\end{figure}

\subsection{Automated Generation and Refinement of Treatment Strategies}
We benchmarked the Strategy Agent against a comprehensive suite of state-of-the-art models, including closed-source leaders (ChatGPT-4o, Claude 3.5 Sonnet) and high-performance open-source models (GLM-4-32B, DeepSeek-V3, DeepSeek-R1, Qwen3-14B, Yi-34B-Chat, Llama-4-Scout, Moonlight-16B, Doubao-1.5-pro). As presented in Table 2, the fine-tuned FUAS-Agents model consistently outperformed all baselines, achieving a ROUGE-L score of 0.4269, which significantly surpassed both GPT-4 (0.2359) and GLM-4-32B (0.1374). This performance gap underscores the limitation of general foundation models in integrating quantitative parameters, such as specific lesion volumes and dose values, into coherent clinical narratives. Qualitative case analysis (Appendix B) further highlighted the system's reasoning capabilities; unlike baseline models such as DeepSeek-R1, which provided generic preoperative checklists but overlooked complex patient-specific contraindications, FUAS-Agents successfully identified high-risk anatomical constraints mentioned in MRI reports. The Optimizer Agent utilized this information to automatically adjust sonication paths and append critical intraoperative warnings, demonstrating a "chain-of-thought" logic that closely mirrors expert-level surgical planning.

\begin{table}[htbp]
\centering
\small 
\caption{Model Performance Comparison}
\label{tab:model_performance}
\setlength{\tabcolsep}{4pt} 
\begin{tabular}{l*{7}{r}} 
\toprule
\multirow{2}{*}{\textbf{Model}} & \multicolumn{3}{c}{\textbf{ROUGE}} & \multicolumn{4}{c}{\textbf{BLEU}} \\
\cmidrule(lr){2-4} \cmidrule(lr){5-8}
& {\textbf{R-1}} & {\textbf{R-2}} & {\textbf{R-L}} & {\textbf{B-1}} & {\textbf{B-2}} & {\textbf{B-3}} & {\textbf{B-4}} \\
\midrule
GPT-4 & 0.3291 & 0.1339 & 0.2359 & 0.3572 & 0.1192 & 0.0551 & 0.0342 \\
ChatGPT-4o & 0.3535 & 0.1121 & 0.1713 & 0.2004 & 0.0660 & 0.0288 & 0.0163 \\
Claude 3 Sonnet & 0.3483 & 0.1004 & 0.1864 & 0.2198 & 0.0628 & 0.0258 & 0.0155 \\
LLaMA 4 Scout 17B & 0.3480 & 0.1159 & 0.1952 & 0.1952 & 0.2717 & 0.0830 & 0.0386 \\
DeepSeek-V3 & 0.3655 & 0.1137 & 0.1927 & 0.2114 & 0.0651 & 0.0291 & 0.0172 \\
DeepSeek-R1 & 0.3368 & 0.0942 & 0.1581 & 0.1672 & 0.0470 & 0.0181 & 0.0094 \\
GLM-4-32B & 0.3234 & 0.0934 & 0.1374 & 0.1219 & 0.0412 & 0.0169 & 0.0087 \\
Moonlight 16B & 0.3419 & 0.1174 & 0.1999 & 0.2431 & 0.0760 & 0.0343 & 0.0208 \\
Yi-34B-Chat & 0.3570 & 0.1130 & 0.1884 & 0.2265 & 0.0697 & 0.0298 & 0.0187 \\
Qwen3-14B & 0.3329 & 0.0823 & 0.1291 & 0.1191 & 0.0329 & 0.0129 & 0.0073 \\
Doubao-1.5-pro & 0.3102 & 0.0725 & 0.1298 & 0.1314 & 0.0344 & 0.0135 & 0.0076 \\
\midrule
\rowcolor{lightgray}
FUAS & \textbf{0.5512} & \textbf{0.3267} & \textbf{0.4269} & \textbf{0.4988} & \textbf{0.2765} & \textbf{0.1806} & \textbf{0.1300} \\
\bottomrule
\end{tabular}
\end{table}

\subsection{Contribution of System Components to Clinical Reliability}
To strictly quantify the specific contribution of each module within the FUAS-Agents architecture, we conducted a systematic ablation study on 20 representative cases (Figure3). The "Full-Function" baseline was compared against three variant configurations. The most critical impact was observed in Ablation Group 1 (No Executor), where the removal of the Executor Agent precipitated a precipitous drop in report Completeness (from 82.5\% to 36.3\%). This finding empirically confirms that without the integration of deterministic imaging tools, the system fails to ground its recommendations in patient-specific anatomical reality, reverting to generic medical advice. In Ablation Group 2 (No Optimizer), the primary degradation was seen in Clinical Compliance, which fell significantly from 97.5\% to 72.5\%. This underscores the indispensability of the self-reflective mechanism; without the Optimizer's safety checks against clinical guidelines, the generation process is prone to hallucinations and regulatory deviations. Finally, Ablation Group 3 (No Memory) demonstrated a marked reduction in Accuracy (from 80.0\% to 58.8\%), validating that access to a retrieval-augmented knowledge base of historical cases is essential for maintaining high-precision diagnostic reasoning.

\begin{figure}[h]
\centering
\includegraphics[width=0.5\textwidth]{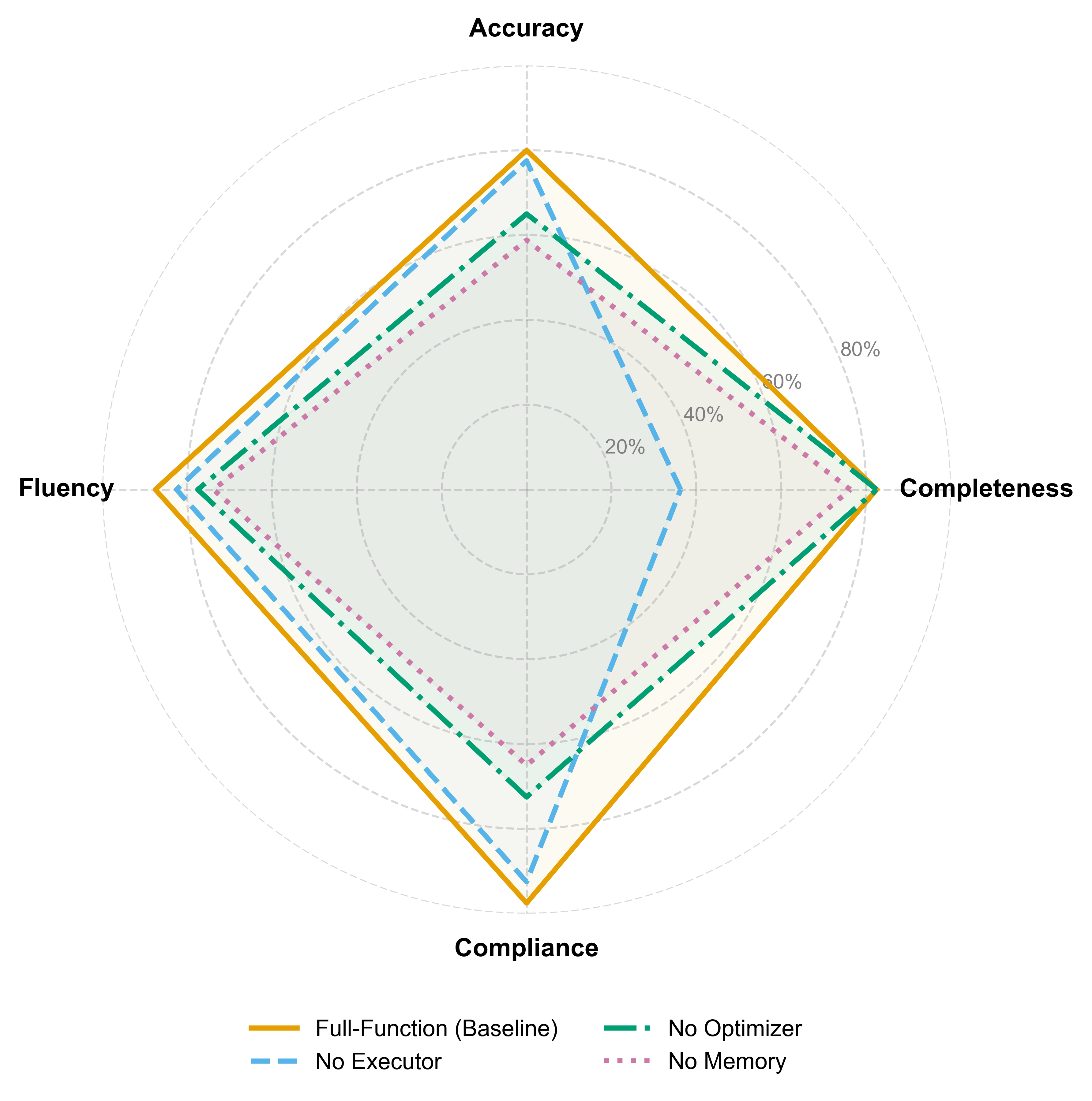}
\caption{Ablation Study Results.For more details, please see Appendix C.}\label{fig:Ablation}
\end{figure}

\subsection{Operational Efficiency Analysis}
We evaluated the computational feasibility of the framework for real-time clinical deployment by analyzing runtime, token usage, and task success rates across all four agents. Detailed performance metrics are documented in Appendix D. The modular design allows for the parallelization of computationally intensive tasks—such as 3D segmentation by the Executor Agent—on dedicated GPU servers, while the reasoning agents (Planner, Strategy, Optimizer) remain lightweight. This architecture ensures that the average end-to-end inference time remains well within the acceptable preoperative planning window. The analysis confirms that FUAS-Agents achieves a high success rate in task execution without introducing prohibitive latency, making it a viable solution for fast-paced surgical environments.

\subsection{Expert Validation and Comparative Assessment}
To assess the translational potential of the system, four senior FUAS specialists independently evaluated 20 randomly selected treatment plans using a standardized 5-point Likert scale (detailed in Appendix E). The system demonstrated high clinical fidelity across all dimensions: 82.5\% of plans were rated as comprehensive in Completeness, and 97.5\% achieved a Compliance score of 4 or higher, indicating strict adherence to ethical and safety standards. In terms of professional presentation, 87.5\% of the reports were deemed fluent and coherent. Furthermore, to benchmark our vertical framework against general medical LLMs, we conducted a comparative evaluation with Baichuan-M2, Huatuo GPT, and LLaVA-Med (Figure 4). Human expert scoring revealed that while general medical models can generate plausible text, they lack the specific domain depth required for FUAS planning. FUAS-Agents consistently outperformed these baselines, highlighting the necessity of a specialized, agent-based architecture for complex surgical decision-making.

\begin{figure}[htbp]
    \centering
    \includegraphics[width=0.6\linewidth]{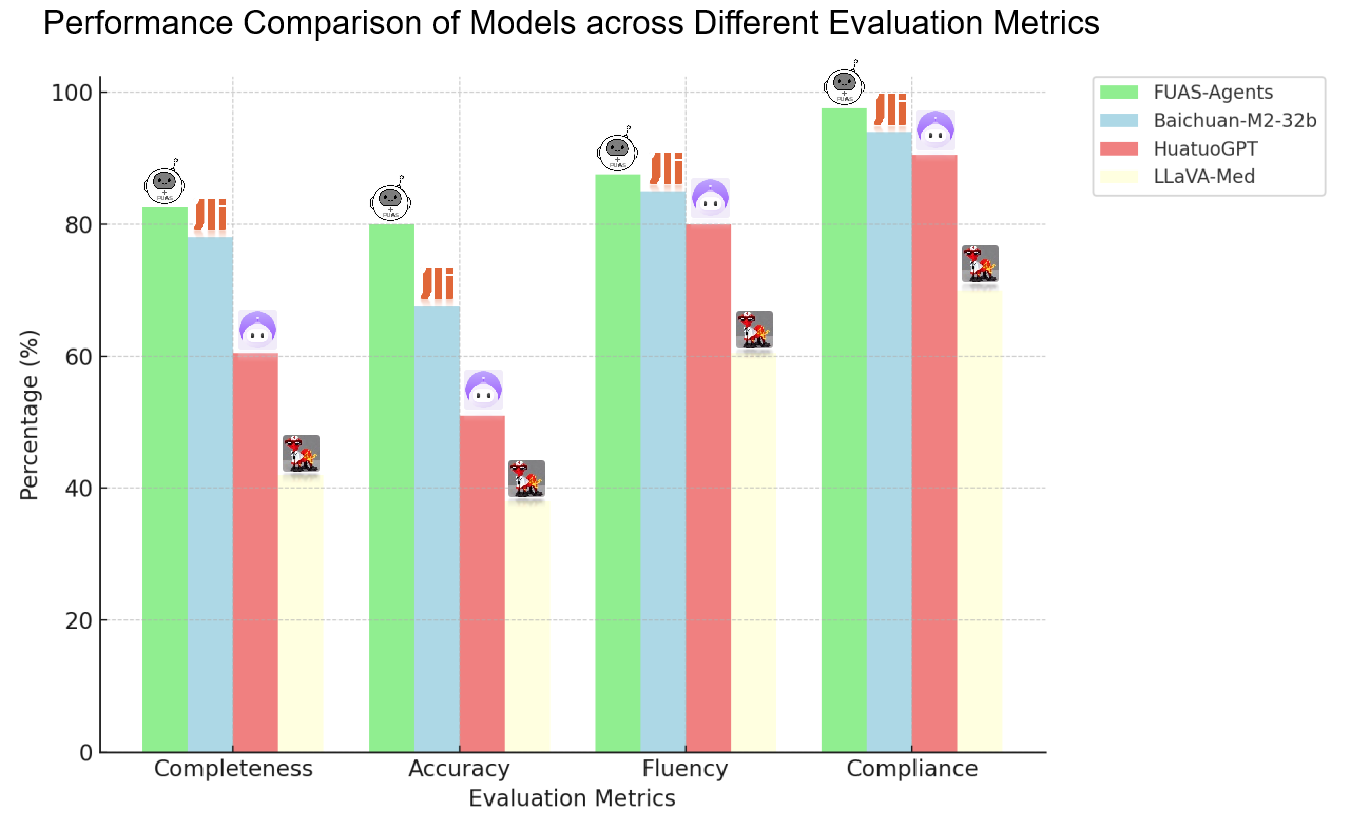}
    \caption{Human evaluation across different models}
    \label{fig:expert_eval}
\end{figure}

\subsection{Case study}
To illustrate the clinical reasoning behavior and strategy adaptation capability of the proposed HIFU intelligent agent, Figure 5 presents four representative uterine fibroid cases, selected to reflect distinct and commonly encountered MRI phenotypes in clinical practice: low T2-signal solitary uterine fibroid: (i) low T2-signal solitary uterine fibroid, (ii) iso-high T2-signal multiple uterine fibroids, (iii) iso–high T2-signal solitary uterine fibroid, and (iv) low T2-signal multiple uterine fibroids. These cases span variations in signal intensity, lesion multiplicity, vascular enhancement patterns, and anatomical location, thereby providing a comprehensive testbed for evaluating agent performance under heterogeneous treatment scenarios.

\begin{figure}[htbp]
    \centering
    
    \begin{subfigure}{0.45\textwidth}
        \centering
        \includegraphics[width=\linewidth]{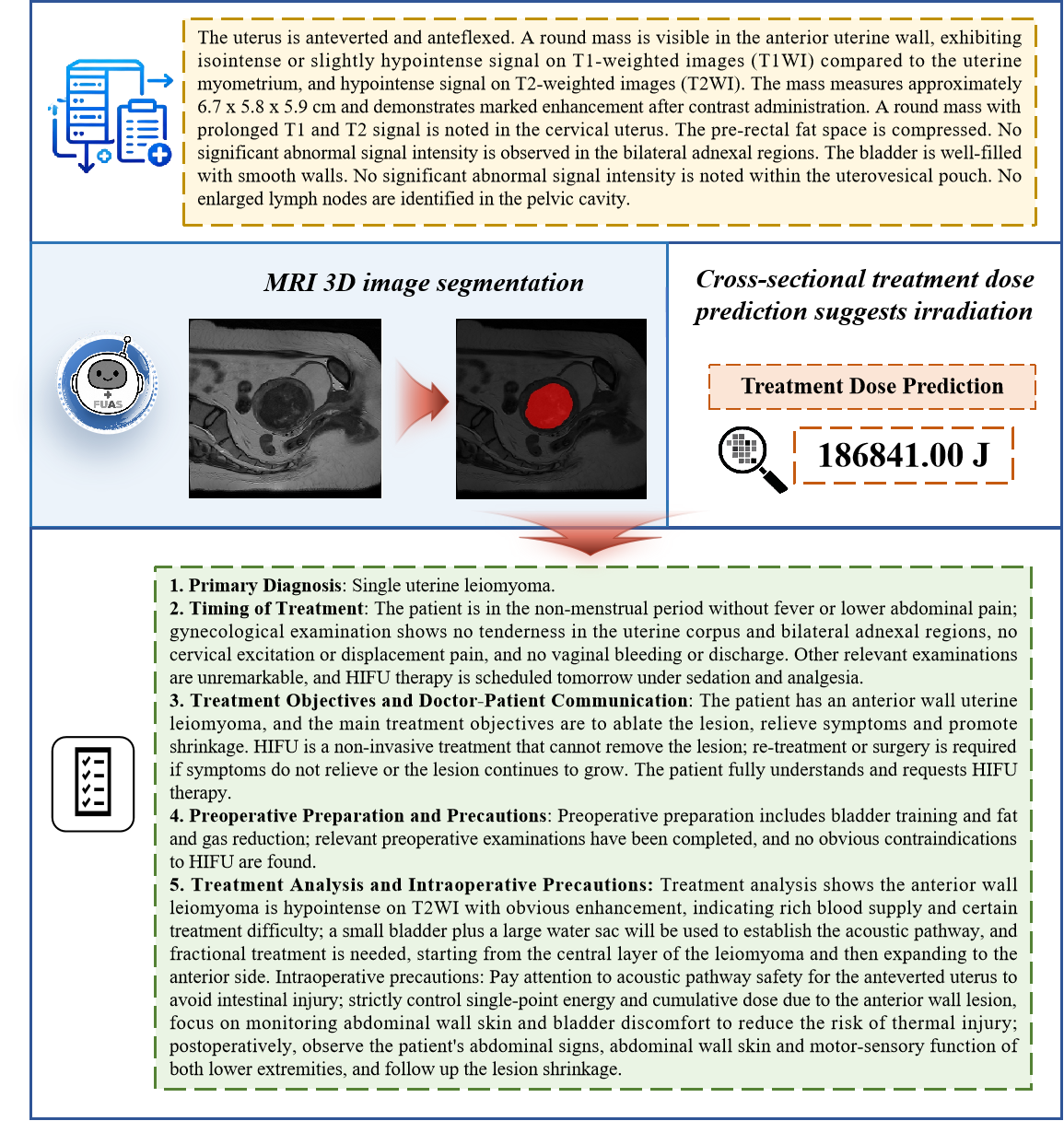}
        \caption{low T2-signal solitary uterine fibroid}
        \label{fig:expert_eval_a}
    \end{subfigure}
    \hfill
    \begin{subfigure}{0.45\textwidth}
        \centering
        \includegraphics[width=\linewidth]{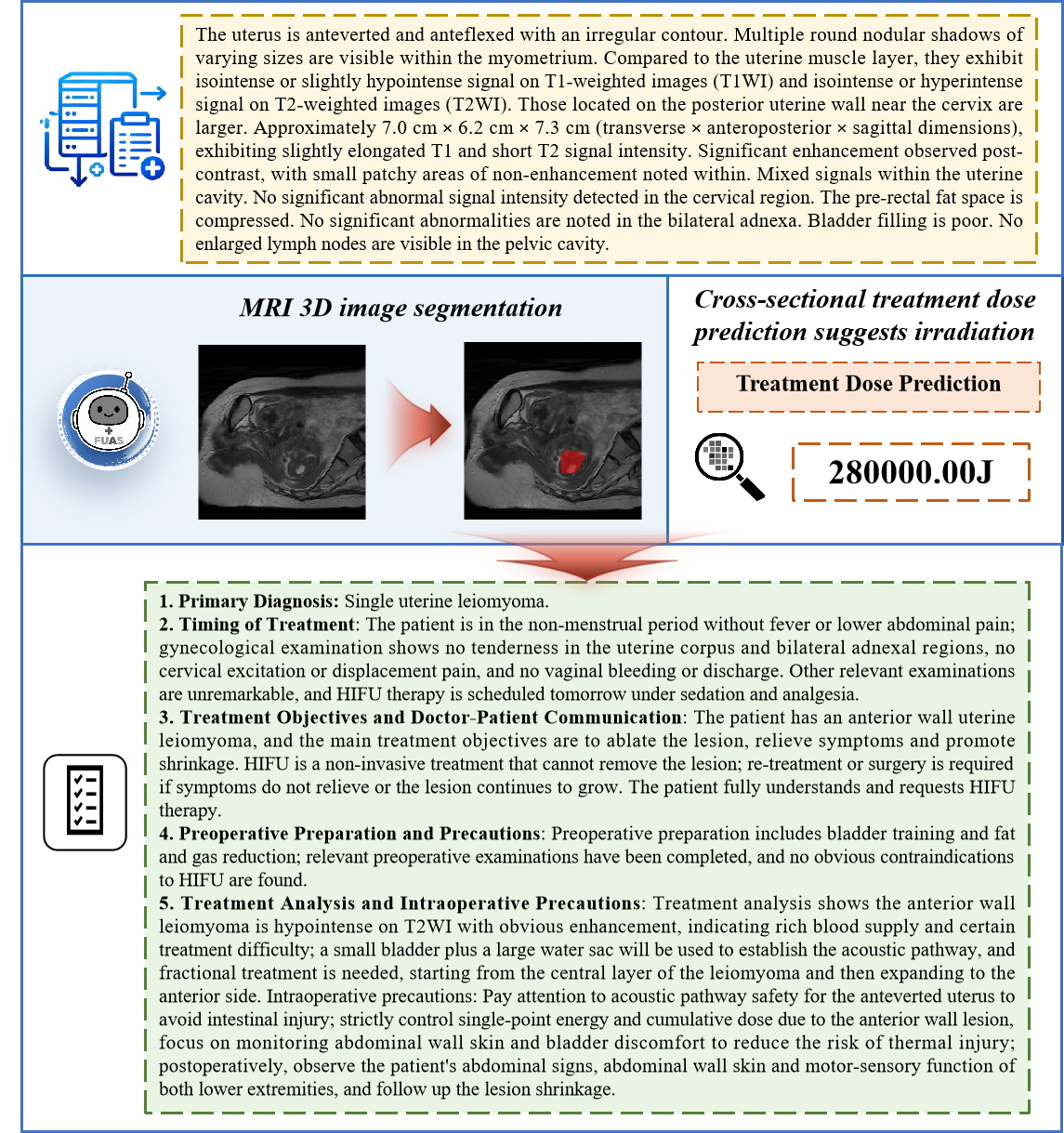}
        \caption{iso-high T2-signal multiple uterine fibroids}
        \label{fig:expert_eval_b}
    \end{subfigure}
    
    \vspace{0.5cm} 
    
    \begin{subfigure}{0.45\textwidth}
        \centering
        \includegraphics[width=\linewidth]{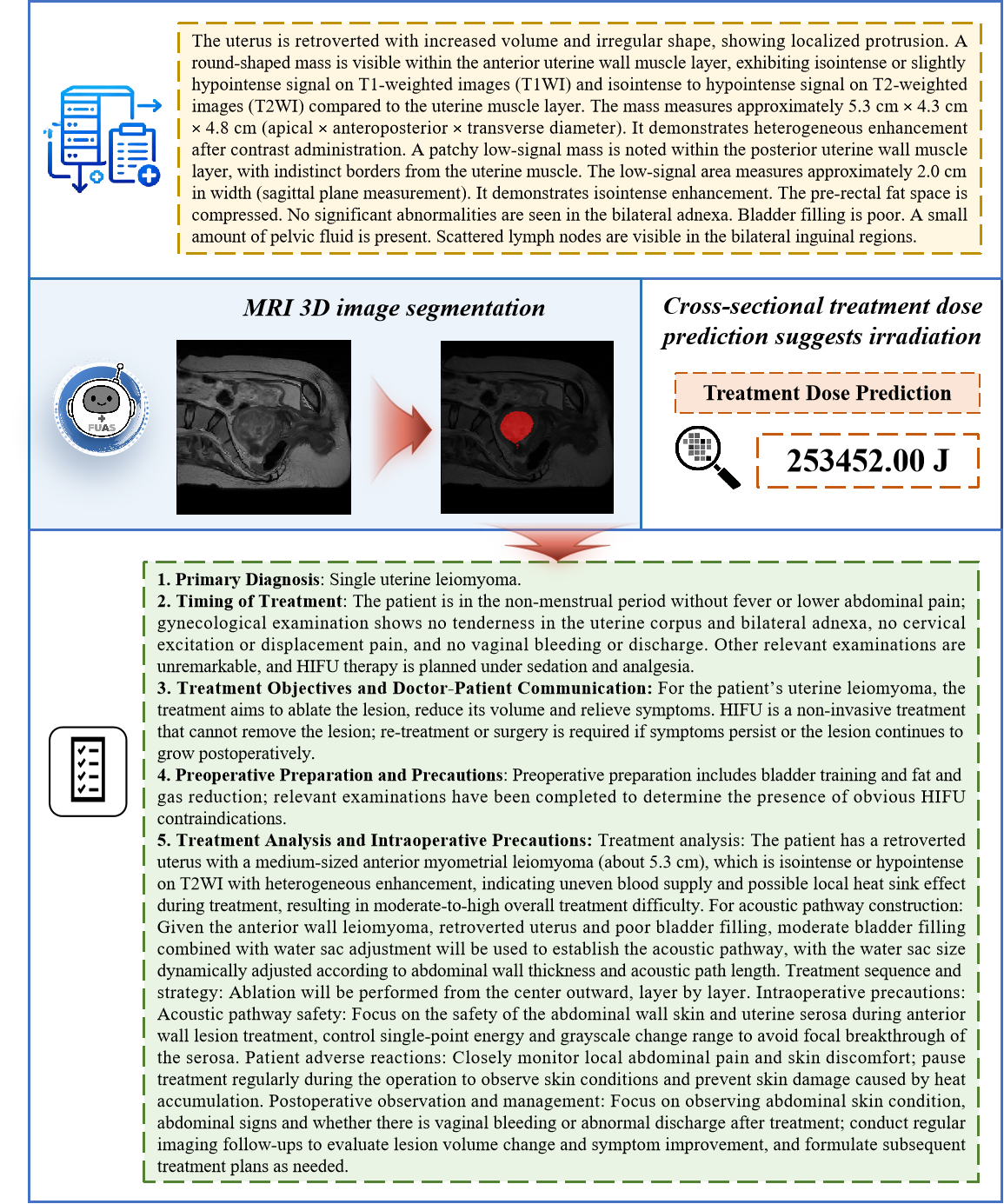}
        \caption{iso-high T2-signal solitary uterine fibroid}
        \label{fig:expert_eval_c}
    \end{subfigure}
    \hfill
    \begin{subfigure}{0.45\textwidth}
        \centering
        \includegraphics[width=\linewidth]{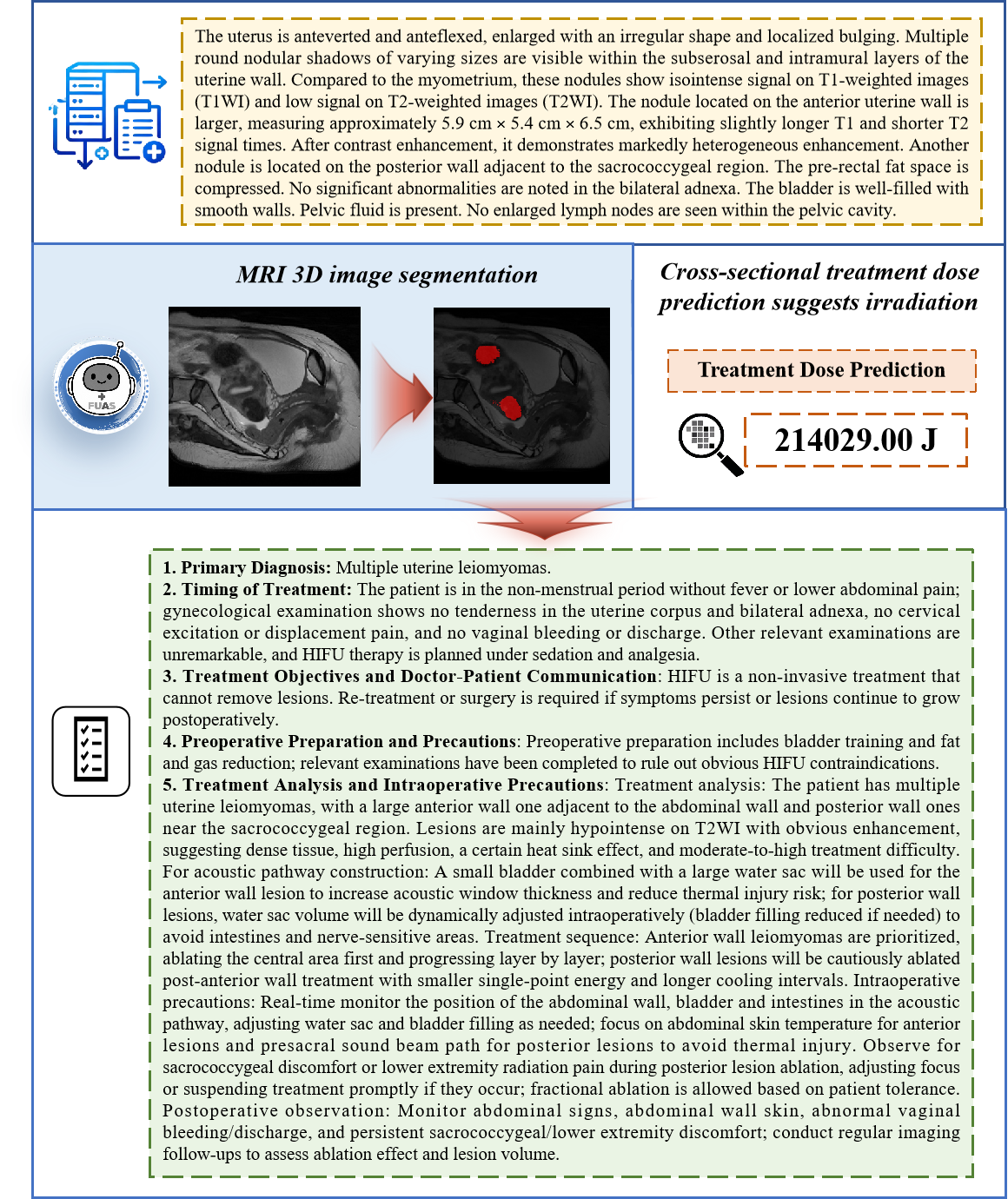}
        \caption{low T2-signal multiple uterine fibroids}
        \label{fig:expert_eval_d}
    \end{subfigure}
    
    \caption{Four representative uterine fibroid cases}
    \label{fig:expert_eval}
\end{figure}
For each case, the agent processes patient-specific MRI data to generate three structured outputs: automated lesion segmentation, cross-sectional treatment dose prediction, and a safety-aware treatment strategy.  In cases of low-complexity, low-signal solitary fibroids, the system recommends a center-to-periphery ablation strategy, utilizing a relatively favorable anterior acoustic window inferred from anatomical structures. When dealing with multiple fibroids, the system clearly distinguishes targets based on acoustic path length, prioritizing lesions behind the uterus while allocating relatively lower energy output and longer cooling intervals to anterior targets, thereby mitigating potential adverse reactions due to heat buildup. Furthermore, in isointense to high-signal cases, the heterogeneity of blood supply reflected in the imaging becomes a significant constraint on decision-making. For multiple high-signal lesions adjacent to the rectum, the system prioritizes strict control of safety boundaries in high-risk scenarios, adopting a phased treatment strategy to balance efficacy and safety. These results demonstrate that FUAS-Agents can generate robust, interpretable, and safety-prioritized treatment strategies, which aligns with expert consensus.

%% file: sections/3discussion.tex
\section{Discussion}

In this study, we propose FUAS-Agents, a multi-agent system designed for treatment planning in FUAS. By integrating multimodal large language models into a clinical decision-making framework, FUAS-Agents enables end-to-end modeling of FUAS treatment planning. Systematic evaluation in the uterine fibroid treatment scenario demonstrates that the proposed system achieves stable and consistent performance across key dimensions, including completeness, accuracy, fluency, and clinical compliance, highlighting its potential utility in complex clinical decision-support tasks.

FUAS-Agents is not intended to replace clinicians by enabling automated decision-making, but rather to function as a clinical decision-support tool for the preoperative planning stage. The treatment plans generated by the system are presented as interpretable and traceable intermediate decision outputs and structured recommendations, with explicit provisions for human review, modification, and rejection, thereby ensuring that final treatment decisions remain informed and clinician-led. In practical clinical settings, FUAS-Agents more closely resembles an auxiliary planning assistant equipped with systematic memory and reasoning capabilities, whose role is to support clinicians in the structured integration of patient-specific information, imaging characteristics, anatomical constraints, and safety regulations, thereby reducing the risk of overlooking critical factors. For FUAS procedures, which are highly experience-dependent and sensitive to safety boundaries, such a human–AI collaborative design paradigm constitutes a key prerequisite for achieving clinical acceptability and sustainable deployment of artificial intelligence systems.

Unlike most existing artificial intelligence approaches that focus on isolated subtasks, FUAS-Agents is developed from the perspective of real-world clinical decision processing and provides a holistic modeling of the multi-stage decision-making process, heterogeneous information sources, and multiple clinical constraints inherent to FUAS treatment. Through explicit division of labor among multiple agents, the system decomposes and organizes task planning, tool execution, strategy generation, memory retrieval, and result optimization in a coordinated manner. This architectural design structurally mirrors the iterative reasoning process of clinicians as they alternate between image interpretation, dose assessment, experience recall, and guideline verification. Such modeling enables the system to better address the experience-intensive and difficult-to-standardize nature of FUAS clinical decision-making.

In comparative experiments, FUAS-Agents consistently outperformed several representative medical large language models in expert-based evaluations. These results indicate that even with specialized medical data training, single-model, language-centric approaches are insufficient for FUAS treatment planning, as FUAS treatment planning is characterized by strong constraints and deep reliance on multimodal information\cite{Vavekanand2026LLMmedical}\cite{Maity2025LLMhealthcare}. Further ablation studies show that removing any key agent module leads to a significant degradation in overall performance, confirming that the observed performance gains arise from coordinated multi-agent collaboration and modular system design rather than from any single component in isolation\cite{Huang2025AgenticAItools}\cite{Pan2024ANSWERED}.

The performance advantages of FUAS-Agents largely stem from its hybrid “foundation model–expert model” collaborative architecture\cite{wang2025perspective}. In this design, multimodal large language models are responsible for task understanding, clinical reasoning, and strategy generation, while critical components with high medical reliability requirements, such as image segmentation and dose prediction, are handled by domain-trained specialized models. This separation of responsibilities expands the system’s reasoning capability while effectively mitigating the controllability and safety risks associated with relying solely on general-purpose models\cite{Buess2025MultimodalAI}\cite{Ferber2025OncologyAgent}\cite{Jenko2025TrustworthyAI}. In addition, the system incorporates a RAG mechanism based on curated clinical knowledge bases and historical case repositories, enabling treatment strategies to explicitly reference established guidelines and real-world precedents, thereby enhancing traceability and consistency in decision-making. To further ensure clinical safety, FUAS-Agents implements two-round quality verification and termination mechanisms: when generated plans fail to satisfy core clinical constraints after repeated optimization, the system proactively halts automated decision-making and triggers human intervention to prevent the accumulation of potential risks\cite{Topol2019HighPerformanceMedicine}\cite{Passerini2025HybridReasoning}.

In the case study, FUAS-Agents demonstrated robust performance across four representative clinical scenarios characterized by substantial heterogeneity in T2WI signal intensity, lesion multiplicity, vascular enhancement patterns, and anatomical location—factors that are closely associated with varying levels of procedural risk and planning complexity in FUAS. Within these heterogeneous settings, FUAS-Agents consistently adapted ablation strategies, acoustic window construction, and safety prioritization, thereby demonstrating its capacity for structured reasoning under diverse and concurrent constraints. Importantly, this capability arises from the explicit modeling of the clinical decision-making process, rather than reliance on predefined treatment templates, suggesting the system’s potential to generalize to previously unseen phenotype combinations and broader clinical planning scenarios.

Several limitations of this study should be acknowledged. Although the effectiveness of the proposed multi-agent system has been demonstrated for FUAS treatment planning, the current design primarily focuses on preoperative planning and decision support, without directly addressing intraoperative real-time control or closed-loop feedback, which involve substantially higher clinical and technical complexity. In addition, some evaluation outcomes rely on expert-based subjective assessments. While multi-dimensional criteria were adopted to mitigate individual bias, the development of more automated and quantitatively grounded evaluation frameworks remains an important direction for future research.

Looking forward, although FUAS-Agents is developed for focused ultrasound ablation, its core design principles exhibit strong potential for transferability. The combination of multi-agent collaboration and the “foundation model–expert model” paradigm can be naturally extended to other interventional or radiation-based therapies that similarly depend on complex imaging data and individualized planning, such as radiotherapy, interventional ablation, and precision surgical planning. More broadly, the proposed framework offers a generalizable design paradigm for domain-specific medical agent systems, emphasizing the deep integration of foundation models with domain knowledge, specialized tools, and clinical standards. Future work will explore the application of this framework to a wider range of clinical tasks and incorporate real-world clinical feedback to continuously refine system performance.

%% file: sections/4method.tex
\section{Method}

\subsection{Overall Framework}

\begin{figure}[ht]
    \centering
    \vspace{-100pt}
    \includegraphics[width=\textwidth,height=0.75\textheight,keepaspectratio]{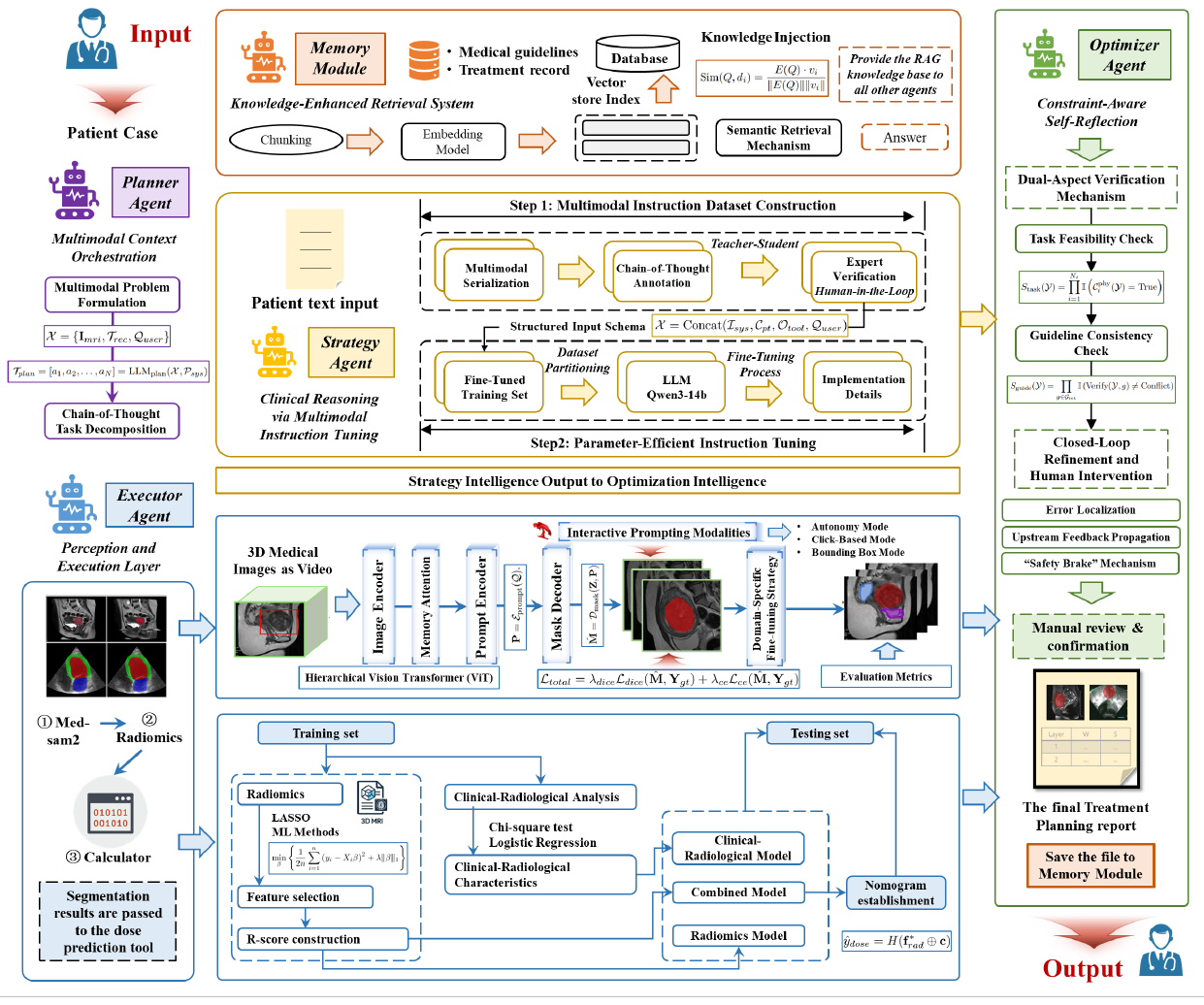}
    \vspace{-90pt}
    \captionsetup{justification=justified, singlelinecheck=false}
    \caption{
        \textbf{System overview of FUAS-Agents.} 
        The framework processes a patient case through a closed-loop multi-agent collaboration. 
        (1) \textbf{Input \& Planning:} The \textit{Planner Agent} parses multimodal inputs (MRI, EHR) into a structured task queue ($\mathcal{T}_{plan}$), underpinned by a centralized \textit{Memory Module} that provides semantic retrieval of guideline-based knowledge to support the entire agent coalition. 
        (2) \textbf{Reasoning Engine:} The \textit{Strategy Agent} serves as the core reasoning unit, powered by a Multimodal Instruction Tuning pipeline. It leverages \textbf{Expert-Guided Data Acquisition} to construct high-fidelity datasets and performs parameter-efficient fine-tuning on Qwen3-14b to enable clinical Chain-of-Thought reasoning. 
        (3) \textbf{Perception \& Quantification:} The \textit{Executor Agent} is deployed for interaction with the physical environment, utilizing a Hierarchical ViT-based MedSAM-2 for 3D lesion segmentation (supporting Autonomy/Click/Box prompting), followed by a Radiomics-LASSO pipeline to predict ablation doses. 
        (4) \textbf{Safety Verification:} The \textit{Optimizer Agent} acts as the final guardrail by enforcing a ``Constraint-Aware Self-Reflection'' mechanism. It conducts dual-aspect checks (Task Feasibility \& Guideline Consistency) to iteratively refine the plan before final output.
    }
    \label{fig:fuas-overview}
\end{figure}

FUAS-Agents is a modular multi-agent framework specifically developed for personalized treatment planning in Focused Ultrasound Ablation Surgery. 
It simulates the clinical decision-making process by comprehensively integrating multimodal medical imaging, structured patient information, and domain-specific clinical knowledge. 
The system follows the ``agent--tool--memory'' paradigm, orchestrating a closed-loop workflow: it progresses sequentially from intent parsing and visual perception to strategy reasoning, while incorporating a safety verification mechanism that triggers iterative refinement upon constraint violation. 

As illustrated in Figure~\ref{fig:fuas-overview}, the architecture consists of three fundamental pillars:

\paragraph{Collaborative Agent Coalition}
The framework delegates decision-making authority to a coalition of four specialized agents, each assuming a distinct cognitive role:
\begin{itemize}
    \item \textit{Planner Agent (Orchestrator):} Acts as the central scheduler, initiating the workflow by parsing unstructured clinical intents and decomposing them into an ordered execution graph of atomic subtasks. 
    \item \textit{Executor Agent (Perception \& Action):} Serves as the sensory interface, bridging the gap between abstract reasoning and physical data by invoking vision tools to segment anatomical structures and quantify ablation doses.
    \item \textit{Strategy Agent (Reasoning Core):} Functions as the primary planner, synthesizing patient-specific treatment protocols by fusing quantitative observations with retrieved clinical evidence.
    \item \textit{Optimizer Agent (Guardrail):} Operates as an inference-time verification module, auditing generated plans against safety constraints and providing feedback to the Planner for re-planning when necessary. 
\end{itemize}

\paragraph{Deterministic Tool Interface}
While LLMs provide powerful semantic reasoning, they lack precision in spatial calculation. To address this, FUAS-Agents integrates a suite of deterministic tools—such as the MedSAM-2 segmentation engine and radiomics physical models. These tools provide the agents with grounded, quantitative capabilities, transforming raw medical imagery into structured features that serve as the factual basis for subsequent strategy generation.

\paragraph{RAG-Enhanced Memory Module}
To mitigate the knowledge cutoff inherent in static models, the system incorporates an external Memory Module. Leveraging a RAG mechanism, this module provides dynamic access to a repository of clinical guidelines, expert consensus documents, and historical case libraries. This ensures that all agent reasoning is theoretically grounded and aligned with the latest medical standards.


\subsection{Agent and Core Module Design}

\subsubsection{Planner Agent: Multimodal Context Orchestration}

The Planner Agent serves as the centralized logic core and entry point of the FUAS-Agents framework. Its primary responsibility is to translate unstructured clinical intents into a structured, executable workflow. Acting as the system's architect, it parses the heterogeneous input data, resolves dependencies between clinical modules, and synthesizes a step-by-step execution graph. This ensures that downstream agents (Executor and Strategy) operate on aligned, context-aware instructions rather than isolated data fragments.

\paragraph{Multimodal Problem Formulation.}
We formally define the input state for a specific patient case as a multimodal tuple $\mathcal{X}$, which aggregates all available pre-operative information:
\begin{equation}
\mathcal{X} = \{ \mathbf{I}_{mri}, \mathcal{T}_{rec}, \mathcal{Q}_{user} \},
\end{equation}
where:
\begin{itemize}
    \item $\mathbf{I}_{mri} \in \mathbb{R}^{H \times W \times D}$ denotes the raw volumetric MRI data;
    \item $\mathcal{T}_{rec}$ represents the patient's textual electronic health records (EHR), containing critical metadata such as symptoms, target lesion characteristics, and contraindications;
    \item $\mathcal{Q}_{user}$ is the natural language instruction provided by the clinician (e.g., \textit{``Plan a treatment for the dominant target lesion while prioritizing safety''}).
\end{itemize}
The Planner Agent first performs Semantic Alignment, utilizing a Vision-Language Model (VLM) interface to ground the textual entities in $\mathcal{T}_{rec}$ (e.g., ``dominant target lesion'') to the visual features in $\mathbf{I}_{mri}$.

\paragraph{Chain-of-Thought Task Decomposition.}
Given the context $\mathcal{X}$, the Planner Agent generates a logical task queue $\mathcal{T}_{plan}$ comprising a sequence of atomic subtasks:
\begin{equation}
\mathcal{T}_{plan} = [a_1, a_2, \dots, a_N] = \text{LLM}_{\text{plan}}(\mathcal{X}, \mathcal{P}_{sys}),
\end{equation}
where $\mathcal{P}_{sys}$ is the system-level meta-prompt defining the clinical workflow logic. Each atomic action $a_i$ is a tuple $\langle \text{Agent}, \text{Tool}, \text{Args} \rangle$, explicitly specifying which agent to invoke and with what parameters.

For a standard FUAS procedure, the Planner typically decomposes the request into three strictly ordered phases:
\begin{enumerate}
    \item \textbf{Perception Phase ($a_1$):} Instructs the Executor Agent to invoke MedSAM-2 for lesion segmentation, passing $\mathbf{I}_{mri}$ to extract the ROI mask $\mathbf{M}$.
    \item \textbf{Quantification Phase ($a_2$):} Directs the Executor Agent to perform dose prediction based on the segmented mask $\mathbf{M}$ and physical device parameters.
    \item \textbf{Reasoning Phase ($a_3$):} Triggers the Strategy Agent to synthesize the final surgical plan $Y$, conditioning on the aggregated outputs from previous steps ($\mathcal{X}, \mathbf{M}, \text{Dose}$).
\end{enumerate}
By explicitly modeling task dependencies (e.g., ensuring segmentation completes before dose calculation), the Planner Agent guarantees the logical consistency and operational feasibility of the entire autonomous pipeline.Definitions of symbols used in formulas and data input/output representations in subsequent text are provided in the Appendix F.

\subsubsection{Executor Agent: Perception and Execution Layer}

The Executor Agent functions as the operational interface of the FUAS-Agents framework, bridging the gap between abstract strategic planning and concrete physical parameters. It acts as a tool dispatcher that translates the Strategy Agent's high-level directives into structured clinical observations. The agent's workflow consists of two critical phases: (1) Visual Perception, enabling precise lesion localization via MRI segmentation; and (2) Quantitative Characterization, mapping morphological features to treatment parameters via dose prediction. By converting high-dimensional imaging data into symbolic states (e.g., masks, radiomics features), the Executor Agent provides the necessary ``grounding'' for the subsequent reasoning modules.

\paragraph{MRI Segmentation: Volumetric Perception via MedSAM-2}

Accurate delineation of the target lesion is the prerequisite for safe ablation. We employ Medical SAM 2 (MedSAM-2)~\cite{bib_medical_sam2} as the core segmentation engine. While the original SAM 2~\cite{bib_sam2} demonstrates powerful generalization in natural images, its performance on low-contrast FUAS target lesions is limited by domain shifts. To address this, we utilize MedSAM-2 as the foundational framework and perform domain-specific adaptation to capture the 3D spatial continuity of medical volumes.

\subparagraph{Network Architecture and Formulation.}
MedSAM-2 follows a promptable encoder-decoder paradigm adapted for volumetric data. Given a 3D MRI volume $\mathbf{X} \in \mathbb{R}^{H \times W \times D}$, the inference process is mathematically formulated as follows:

\begin{itemize}
    \item \textbf{Image Encoder ($\mathcal{E}_{\mathrm{img}}$):} A hierarchical Vision Transformer (ViT) processes the input volume to map voxel-level information into a high-dimensional latent representation:
    \begin{equation}
    \mathbf{Z} = \mathcal{E}_{\mathrm{img}}(\mathbf{X}), \quad \mathbf{Z} \in \mathbb{R}^{h \times w \times d \times C}.
    \end{equation}
    Unlike standard 2D encoders, $\mathcal{E}_{\mathrm{img}}$ incorporates a Memory Attention mechanism. It treats the 3D volume as a sequence of slices, aggregating features from adjacent frames to enforce global geometric consistency along the Z-axis.

    \item \textbf{Prompt Encoder ($\mathcal{E}_{\mathrm{prompt}}$):} To incorporate clinical guidance, the model accepts a set of geometric prompts $\mathcal{Q}$ (e.g., points, bounding boxes). These are encoded into sparse embedding vectors:
    \begin{equation}
    \mathbf{P} = \mathcal{E}_{\mathrm{prompt}}(\mathcal{Q}).
    \end{equation}

    \item \textbf{Mask Decoder ($\mathcal{D}_{\mathrm{mask}}$):} The decoder acts as a lightweight fusion module. It integrates image embeddings $\mathbf{Z}$ and prompt embeddings $\mathbf{P}$ via cross-attention to generate the voxel-wise probability map:
    \begin{equation}
    \hat{\mathbf{M}} = \mathcal{D}_{\mathrm{mask}}(\mathbf{Z}, \mathbf{P}),
    \end{equation}
    where $\hat{\mathbf{M}} \in [0,1]^{H \times W \times D}$ represents the predicted segmentation mask.
\end{itemize}

\subparagraph{Domain-Specific Fine-tuning Strategy.}
To bridge the domain gap between natural images and medical MRI, we conducted supervised fine-tuning using our collected dataset of expert-annotated FUAS cases covering diverse target lesion presentations. The training objective is to minimize the discrepancy between the predicted mask $\hat{\mathbf{M}}$ and the ground truth mask $\mathbf{Y}_{gt}$. We employ a composite loss function $\mathcal{L}_{total}$ combining Dice Loss ($\mathcal{L}_{dice}$) and Cross-Entropy Loss ($\mathcal{L}_{ce}$) to handle class imbalance and maintain boundary precision:
\begin{equation}
\mathcal{L}_{total} = \lambda_{dice} \mathcal{L}_{dice}(\hat{\mathbf{M}}, \mathbf{Y}_{gt}) + \lambda_{ce} \mathcal{L}_{ce}(\hat{\mathbf{M}}, \mathbf{Y}_{gt}),
\end{equation}
where $\lambda_{dice}$ and $\lambda_{ce}$ are weighting coefficients. This fine-tuning process aligns the model's feature space with the characteristic texture patterns of FUAS target lesions.

\subparagraph{Interactive Prompting Modalities.}
Leveraging the fine-tuned model, the Executor Agent supports three operation modes to adapt to different clinical workflows:

\begin{enumerate}
    \item \textbf{Autonomy Mode (Zero-Shot):} 
    The agent utilizes an integrated object detection head to automatically generate candidate bounding boxes $\mathcal{Q}_{auto}$ directly from the embedding $\mathbf{Z}$. This enables fully automated lesion screening without human intervention.
    
    \item \textbf{Click-Based Mode (Human-in-the-Loop):} 
    For high-precision requirements, the model accepts sparse point prompts $\mathcal{Q}_{click} = \{(x_i, y_i, z_i, l_i)\}$, where clinicians provide positive (lesion) or negative (background) clicks to iteratively refine the boundary. This mode serves as the primary tool for detailed surgical planning.
    
    \item \textbf{Bounding Box Mode:} 
    Clinicians specify a spatial region of interest $\mathcal{Q}_{bbox}$ via diagonal coordinates. This constrains the segmentation attention to the target lesion, effectively filtering out interference from surrounding organs.
\end{enumerate}

\subparagraph{Evaluation Metrics.}
To quantitatively assess the performance of the Executor Agent, we employ standard overlap-based metrics including the Dice Similarity Coefficient (DSC) and Intersection over Union (IoU). These metrics measure the spatial alignment between the agent's predictions and expert annotations, verifying the system's reliability for clinical deployment.

\paragraph{Dose Prediction: Radiomics-Driven Quantitative Estimation}

The Dose Prediction module serves as a critical quantitative tool within the Executor Agent, bridging the gap between morphological imaging features and physical treatment parameters. Unlike generic regression tasks, this module is designed to model the non-linear mapping from the high-dimensional radiomics feature space to the target energy deposition value required for effective ablation.

\subparagraph{Radiomics Feature Extraction Pipeline.}
Given the lesion mask $\hat{\mathbf{M}}$ generated by the Segmentation module and the original MRI volume $\mathbf{X}$, we first define the Region of Interest (ROI). To capture the heterogeneity of the lesion tissue, we extract a comprehensive set of radiomics features using the PyRadiomics engine.
Let $\mathbf{f}_{rad} \in \mathbb{R}^{d}$ denote the extracted feature vector, which encompasses:
\begin{itemize}
    \item \textit{First-order Statistics:} Characterizing the global intensity distribution (e.g., mean, skewness, entropy).
    \item \textit{Shape Features:} Describing the 3D geometric properties (e.g., sphericity, surface-area-to-volume ratio).
    \item \textit{Texture Features:} Quantifying spatial patterns using Gray Level Co-occurrence Matrix (GLCM) and Gray Level Size Zone Matrix (GLSZM), which serve as proxies for tissue density and vascularization levels.
\end{itemize}

\subparagraph{Sparse Feature Selection.}
The initial feature space is high-dimensional and prone to redundancy. To construct a robust predictor, we employ a two-stage dimensionality reduction strategy. First, features with low inter-observer reproducibility (Intraclass Correlation Coefficient $< 0.75$) are discarded. Second, we apply the Least Absolute Shrinkage and Selection Operator (LASSO) to identify the most informative subset.
The objective is to solve:

\begin{equation}
\min_{\mathbf{w}} \left( \frac{1}{2N} \| \mathbf{y} - \mathbf{F}\mathbf{w} \|_2^2 + \lambda \| \mathbf{w} \|_1 \right),
\end{equation}
where $\mathbf{F}$ is the feature matrix and $\lambda$ controls the sparsity. This process yields a compact feature signature $\mathbf{f}^*_{rad}$ that correlates most strongly with the required sonication energy.

\subparagraph{Heterogeneous Data Fusion and Inference.}
Effective dose prediction requires context beyond imaging. We introduce a feature fusion mechanism that integrates the radiomics signature $\mathbf{f}^*_{rad}$ with structured clinical variables $\mathbf{c}$ (e.g., patient demographics, acoustic pathway features such as intervening tissue thickness, and target anatomical location). The final predictor $H(\cdot)$ is implemented as an XGBoost regressor, chosen for its ability to handle non-linear interactions between heterogeneous modalities:

\begin{equation}
\hat{y}_{dose} = H(\mathbf{f}^*_{rad} \oplus \mathbf{c}),
\end{equation}

where $\oplus$ denotes feature concatenation. The model is trained to minimize the Root Mean Squared Error (RMSE) between the predicted energy and the actual successful ablation dose recorded in historical logs.
The output $\hat{y}_{dose}$ is subsequently passed to the Strategy Agent as a quantitative prior ($F_T$) to guide the generation of the comprehensive treatment plan.For the dose prediction module in FUAS-Agents. A detailed comparison of different machinelearning models is provided in Appendix G.

\subsubsection{Strategy Agent: Clinical Reasoning via Multimodal Instruction Tuning}

The Strategy Agent serves as the cognitive engine of the FUAS-Agents framework. Unlike traditional end-to-end models that function as "black boxes," this agent operates on a robust ``perceive-reason-act'' loop. It synthesizes structured patient records with quantitative tool observations (e.g., segmentation masks, dose metrics) to generate evidence-based treatment plans~\cite{bib_psydt}. The methodological construction of this agent involves two critical phases: the curation of a domain-specific multimodal instruction dataset and the application of parameter-efficient fine-tuning.

\paragraph{Multimodal Instruction Dataset Construction.}
To bridge the gap between generic LLM reasoning and specific FUAS protocols, we constructed a high-fidelity instruction-following dataset, denoted as $\mathcal{D}_{instruct}$.
Let $\mathcal{D}_{raw}$ represent the raw retrospective cohort comprising over 2,000 historical cases. We transform these unstructured records into structured training pairs $\{( \mathcal{X}_i, \mathcal{Y}_i )\}$ through a three-step pipeline:

\begin{enumerate}
    \item \textit{Multimodal Serialization:} 
    Since standard LLMs cannot process volumetric MRI data directly, we utilize the Executor Agent as a modality bridge. The visual segmentation results are serialized into geometric descriptors $\mathcal{O}_{seg}$ (e.g., lesion volume, spatial coordinates, adjacency to organs at risk). Similarly, dose predictions are converted into quantitative tokens $\mathcal{O}_{dose}$. These observational tokens are concatenated with the textual patient profile $\mathcal{C}_{pt}$ (demographics and history) to form the unified input context $\mathcal{X}$.
    
    \item \textit{Chain-of-Thought (CoT) Annotation:} 
    A simple input-output mapping is insufficient for learning complex clinical logic. We employ a "Teacher-Student" paradigm, using GPT-4 guided by rigorous clinical guidelines to generate an intermediate reasoning trace $\mathcal{R}$. This trace explicitly articulates \textit{why} a specific plan is chosen based on $\mathcal{X}$. Consequently, the target output is formulated as a composite sequence $\mathcal{Y} = [\mathcal{R} ; \mathcal{Y}_{plan}]$, where $;$ denotes sequence concatenation.
    
    \item \textit{Expert Verification:} 
    To ensure medical validity, a Human-in-the-Loop (HITL) protocol was implemented. Five senior FUAS specialists reviewed and refined the generated reasoning paths, ensuring the ground truth aligns with the latest consensus~\cite{bib_zodiac}.
\end{enumerate}

\paragraph{Structured Input Schema.}
Instead of raw text concatenation, we define the input prompt $\mathcal{X}$ as a hierarchical structured sequence to ensure semantic clarity:
\begin{equation}
\mathcal{X} = \text{Concat}(\mathcal{I}_{sys}, \mathcal{C}_{pt}, \mathcal{O}_{tool}, \mathcal{Q}_{user}),
\end{equation}
where $\mathcal{I}_{sys}$ represents the system-level instruction defining the agent's role as an FUAS expert; $\mathcal{C}_{pt}$ denotes the structured patient profile; $\mathcal{O}_{tool} = \{\mathcal{O}_{seg}, \mathcal{O}_{dose}\}$ aggregates the serialized quantitative findings from the Executor Agent; and $\mathcal{Q}_{user}$ is the specific planning instruction.

\paragraph{Parameter-Efficient Instruction Tuning.}
We selected Qwen3-14B as the foundational model due to its superior reasoning benchmarks. To adapt this general-purpose model to the specialized FUAS domain while maintaining computational efficiency, we employ Low-Rank Adaptation (LoRA).

Formally, let $\Theta_{pre}$ represent the frozen pre-trained weights of the transformer backbone. We introduce trainable low-rank matrices $\mathbf{A}$ and $\mathbf{B}$ to approximate the weight updates $\Delta \mathbf{W}$ in the attention layers. The forward pass for a hidden state $h$ is formulated as:
\begin{equation}
h = \mathbf{W}_0 x + \Delta \mathbf{W} x = \mathbf{W}_0 x + \frac{\alpha}{r} \mathbf{B} \mathbf{A} x,
\end{equation}
where $r=8$ is the rank, and $\alpha=16$ is the scaling factor. The training objective is to minimize the negative log-likelihood of the target sequence $\mathcal{Y}$ conditioned on the multimodal input $\mathcal{X}$:
\begin{equation}
\mathcal{L}(\Theta_{LoRA}) = - \mathbb{E}_{(\mathcal{X},\mathcal{Y}) \in \mathcal{D}_{instruct}} \left[ \sum_{t=1}^{|\mathcal{Y}|} \log P(y_t \mid y_{<t}, \mathcal{X}; \Theta_{pre}, \Theta_{LoRA}) \right].
\end{equation}
By optimizing this objective, the Strategy Agent learns to map the serialized tool observations $\mathcal{O}_{tool}$ to clinically valid reasoning paths and treatment strategies.

\paragraph{Implementation Details.}
The model was fine-tuned for 3 epochs using a global batch size of 8, distributed across four NVIDIA A800 (80GB) GPUs. We utilized the AdamW optimizer with a learning rate of $5 \times 10^{-5}$ and a cosine decay scheduler. To optimize memory footprint without compromising numerical stability, BF16 precision was employed throughout the training phase.

\subsubsection{Optimizer Agent: Constraint-Aware Self-Reflection}

The Optimizer Agent functions as the definitive clinical safety guardrail within the FUAS-Agents framework. Analogous to a senior physician reviewing a junior resident's proposal, it operates on a hierarchical oversight mechanism designed to mitigate the inherent probabilistic hallucinations of Large Language Models. Unlike the Strategy Agent, which focuses on creative solution generation, the Optimizer Agent employs a deterministic ``generate-evaluate-refine'' loop. Its core objective is to ensure that candidate plans are strictly compliant with both physical limitations and dynamic clinical guidelines.

\paragraph{Dual-Aspect Verification Mechanism.}
Upon receiving a candidate treatment plan $\mathcal{Y}$ and the patient context $\mathcal{X}$, the Optimizer Agent initiates a rigorous verification process. This is modeled as a binary scoring function $S(\mathcal{Y}) \in \{0, 1\}$, decomposed into two orthogonal dimensions:

\begin{itemize}
    \item \textit{Task Feasibility Check ($S_{\mathrm{task}}$):} This module enforces hard, static physical constraints (e.g., device power limits, geometric coverage). It is formulated as a conjunction of boolean logic checks:
    \begin{equation}
    S_{\mathrm{task}}(\mathcal{Y}) = \prod_{i=1}^{N_t} \mathbb{I}\left( \mathcal{C}_i^{\mathrm{phy}}(\mathcal{Y}) = \text{True} \right),
    \end{equation}
    where $N_t$ is the number of hard rules, and $\mathbb{I}(\cdot)$ is the indicator function. If any physical constraint $\mathcal{C}_i^{\mathrm{phy}}$ is violated (e.g., predicted dose $E_{pred} > P_{max}$), the entire score becomes 0.

    \item \textit{Guideline Consistency Check ($S_{\mathrm{guide}}$):} Unlike static physical rules, clinical guidelines are context-dependent. We incorporate the RAG mechanism directly into the verification logic.
    First, the agent retrieves relevant guideline fragments $\mathcal{G}_{\mathrm{ret}}$ from the Memory Module ($\mathcal{M}$) based on the plan content:
    \begin{equation}
    \mathcal{G}_{\mathrm{ret}} = \text{Retrieve}(\mathcal{Y}, \mathcal{X}; \mathcal{M}).
    \end{equation}
    Then, the consistency is evaluated by checking the plan against these retrieved protocols:
    \begin{equation}
    S_{\mathrm{guide}}(\mathcal{Y}) = \prod_{g \in \mathcal{G}_{\mathrm{ret}}} \mathbb{I}\left( \text{Verify}(\mathcal{Y}, g) \neq \text{Conflict} \right).
    \end{equation}
    Here, $\text{Verify}(\cdot)$ employs a logical inference step to detect semantic contradictions (e.g., checking if the patient's position in $\mathcal{Y}$ violates the contraindications listed in $g$).
\end{itemize}

\paragraph{Closed-Loop Refinement and Human Intervention.}
The optimization process is formulated as an iterative trajectory. Let $t$ denote the current iteration step. If the composite validity score indicates a failure (i.e., $S_{\mathrm{total}} = S_{\mathrm{task}} \cdot S_{\mathrm{guide}} = 0$), the system triggers a feedback loop governed by the following logic:

\begin{enumerate}
    \item \textit{Error Localization:} The agent isolates the specific set of violated constraints $\mathcal{V}$ and constructs a structured error report $\mathcal{F}_{fb}$ (e.g., \textit{``Violation of Guideline G4: Safety margin < 10mm around critical structures''}).
    
    \item \textit{Upstream Feedback Propagation:} Instead of locally patching the text, the error report $\mathcal{F}_{fb}$ is propagated back to the Planner Agent. This forces the orchestrator to re-evaluate the task context and re-initiate the generation pipeline, ensuring that the correction is architecturally consistent rather than just superficially text-edited.
    
    \item \textit{The "Safety Brake" Mechanism:} To strictly guarantee patient safety and prevent infinite regression loops, we implement a maximum iteration threshold ($T_{max}=2$). The workflow logic is defined as:
    \begin{equation}
    \text{Next Action} = 
    \begin{cases} 
    \text{Trigger Planner w/ } \mathcal{F}_{fb}, & \text{if } S_{\mathrm{total}}=0 \text{ and } t < T_{max} \\
    \text{Escalate to Human Intervention}, & \text{if } S_{\mathrm{total}}=0 \text{ and } t \ge T_{max} 
    \end{cases}
    \end{equation}
    If the problem remains unresolved after two refinement cycles, the system halts autonomous execution and flags the case for manual review by a senior physician, adhering to the "Human-in-the-Loop" safety protocol.
\end{enumerate}

\subsubsection{Memory Module: Knowledge-Enhanced Retrieval System}

The Memory Module ($\mathcal{M}$) serves as the long-term knowledge repository for the FUAS-Agents framework. It is designed to overcome the ``parametric knowledge cutoff'' of standard LLMs by providing up-to-date, domain-specific medical evidence. This module underpins the RAG mechanism utilized by both the Strategy Agent (for informed planning) and the Optimizer Agent (for constraint verification).

\paragraph{Knowledge Base Construction.}
We construct a dedicated clinical vector database derived from authoritative sources, comprising: 
(1) International Clinical Practice Guidelines for FUAS (e.g., ISFUS guidelines); 
(2) A retrospective repository of high-quality historical clinical cases validated by domain experts, serving as successful reference templates; 
and (3) A curated database of contraindications.

The raw textual data is processed via a three-stage pipeline:
\begin{enumerate}
    \item \textbf{Chunking:} Long documents and case logs are segmented into discrete semantic passages $D = \{d_1, d_2, \dots, d_N\}$ with a fixed window size of 512 tokens and a 50-token overlap to preserve context continuity.
    \item \textbf{Embedding:} Each text chunk $d_i$ is transformed into a high-dimensional vector representation $v_i \in \mathbb{R}^{d}$ using the OpenAI \texttt{text-embedding-ada-002} model ($d=1536$). This model captures the semantic essence of medical terminology effectively.
    \item \textbf{Indexing:} The resulting vectors are stored in a dense vector index utilizing FAISS (Facebook AI Similarity Search) to enable millisecond-level retrieval latency.
\end{enumerate}

\paragraph{Semantic Retrieval Mechanism.}
We formally define the retrieval function $\text{Retrieve}(Q; \mathcal{M})$ introduced in the Optimizer Agent section. Given a query $Q$ (typically the patient context $X$ or a generated plan $Y$), the system computes the semantic similarity between the query embedding $E(Q)$ and stored knowledge vectors $v_i$.
We employ Cosine Similarity as the metric:

\begin{equation}
\text{Sim}(Q, d_i) = \frac{E(Q) \cdot v_i}{\|E(Q)\| \|v_i\|}.
\end{equation}

The system retrieves the top-$K$ most relevant chunks $\mathcal{G}_{ret} = \{d_{(1)}, \dots, d_{(K)}\}$ ranked by similarity scores. In our experiments, we set $K=3$ to balance information density and context window usage.

\paragraph{Knowledge Injection.}
The retrieved content $\mathcal{G}_{ret}$ serves two distinct purposes:
\begin{itemize}
    \item \textbf{Reference for Generation:} For the Strategy Agent, retrieved historical cases act as ``few-shot'' demonstrators. By analyzing how similar patients were successfully treated in the past, the model can mimic proven clinical reasoning patterns.
    \item \textbf{Standard for Verification:} For the Optimizer Agent, retrieved guidelines serve as the ground truth for the $S_{\mathrm{guide}}$ consistency check, ensuring the new plan does not violate established safety protocols.
\end{itemize}
By decoupling knowledge storage from model parameters, the Memory Module ensures that the system's clinical knowledge remains explicit, verifiable, and easily updateable without retraining.

\section*{Author contribution}

L.Z., Z.B., and Q.C. drafted the main manuscript text and contributed to the conceptual development of the study. Y.L. and Z.L. prepared the figures and were responsible for graphical editing and formatting. J.B. and G.L. managed the clinical database and supported data organization and preprocessing. M.H. provided clinical expertise and guidance throughout the study. K.L. and H.Y. contributed to methodological supervision, critical revisions, and overall study guidance. Z.Z. conceived the project, supervised the research. All authors reviewed and approved the final manuscript.

\section*{Acknowledgements}

This research was funded by the Tsinghua University Hospital Management Research and Development Project (Project No.: 100011005).

\section*{Conflict of interest}

The authors have no competing interests to declare.

\section*{Ethics Statement}

This study was conducted based on retrospective real-world clinical data from patients who underwent FUAS. The study protocol was approved by the relevant institutional review board (IRB). All data were anonymized and de-identified prior to use and did not contain any personally identifiable information.

\section*{Data Availability}

The data used in this study were collected from clinical data from multiple medical institutions, including sensitive patient information. Due to ethical constraints and data privacy agreements, this data is not publicly available. If there is a reasonable request, an application can be made to the corresponding author, but approval from the relevant institutional review board and a signed data sharing agreement are required.

%% file: sections/5appendix.tex
\clearpage

\section{Segmentation Samples}
\label{appendix:A}

\renewcommand{\thefigure}{A\arabic{figure}}
\setcounter{figure}{0}

\vspace{4cm} 
\begin{figure}[h]
    \centering
    \includegraphics[width=1.0\textwidth]{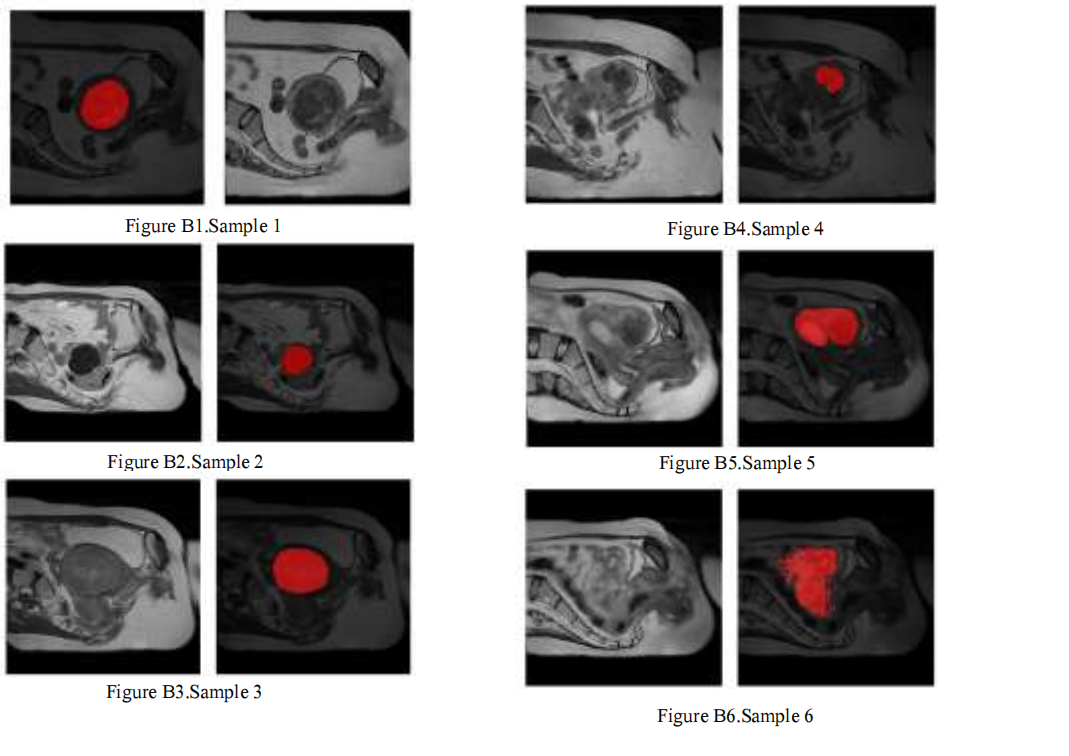}
    \caption{Detailed visual comparisons and segmentation metrics}
    \label{fig:appendixA_1}
\end{figure}
\clearpage

\clearpage

\section{Case Study of Treatment Planning Generation by Different LLMs}
\label{appendix:B}

\renewcommand{\thefigure}{B\arabic{figure}}
\setcounter{figure}{0}

\vspace{4cm} 
\begin{figure}[h]
    \centering
    \includegraphics[width=1.0\textwidth]{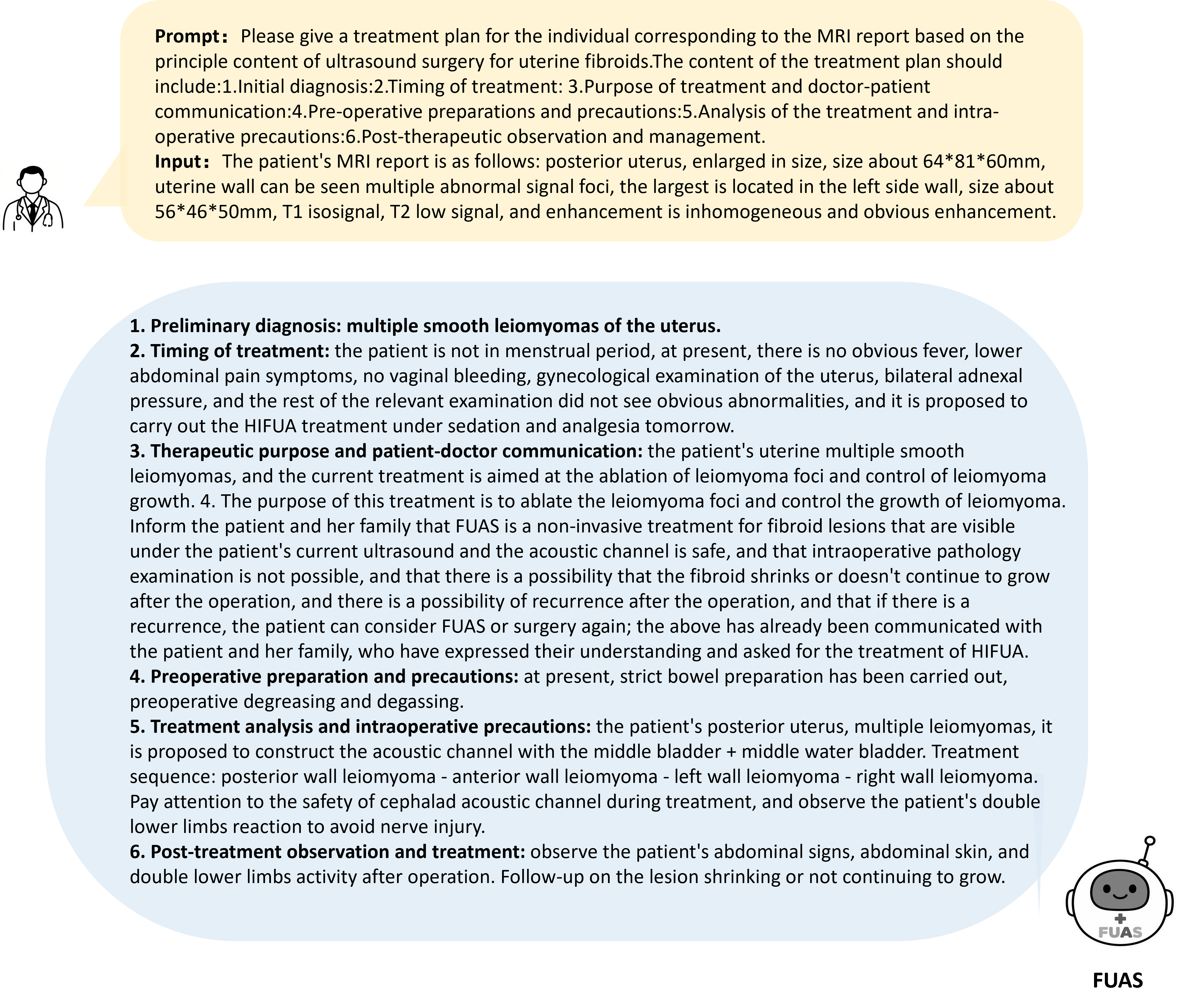}
    \caption{Treatment Plan Generated by FUAS-Agents}
    \label{fig:appendixB_1}
\end{figure}

\begin{figure}[p]
    \centering
    \includegraphics[width=1.0\textwidth]{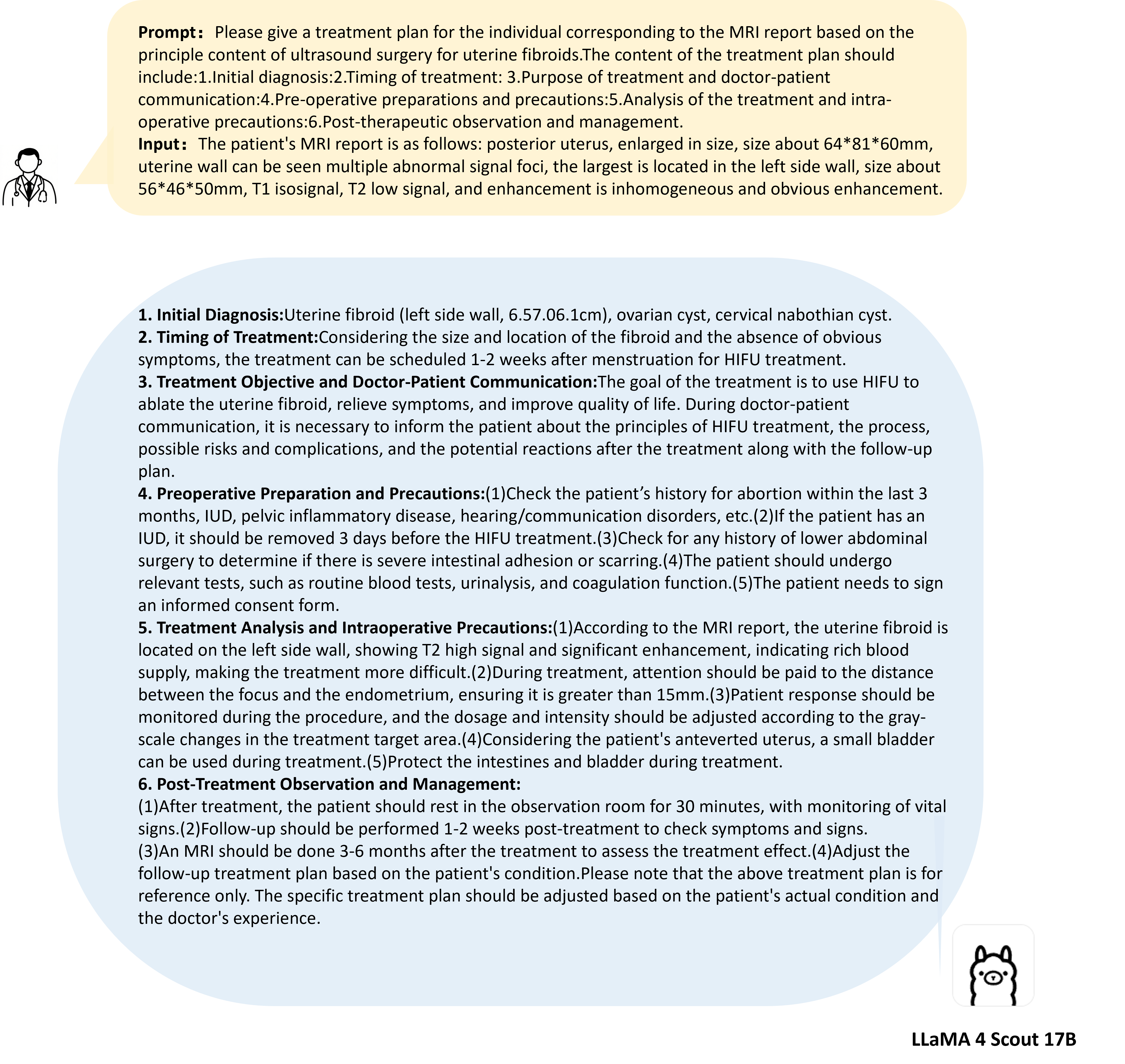}
    \caption{Treatment Plan Generated by LLaMA}
    \label{fig:appendixB_2}
\end{figure}
\clearpage

\begin{figure}[p]
    \centering
    \includegraphics[width=1.0\textwidth]{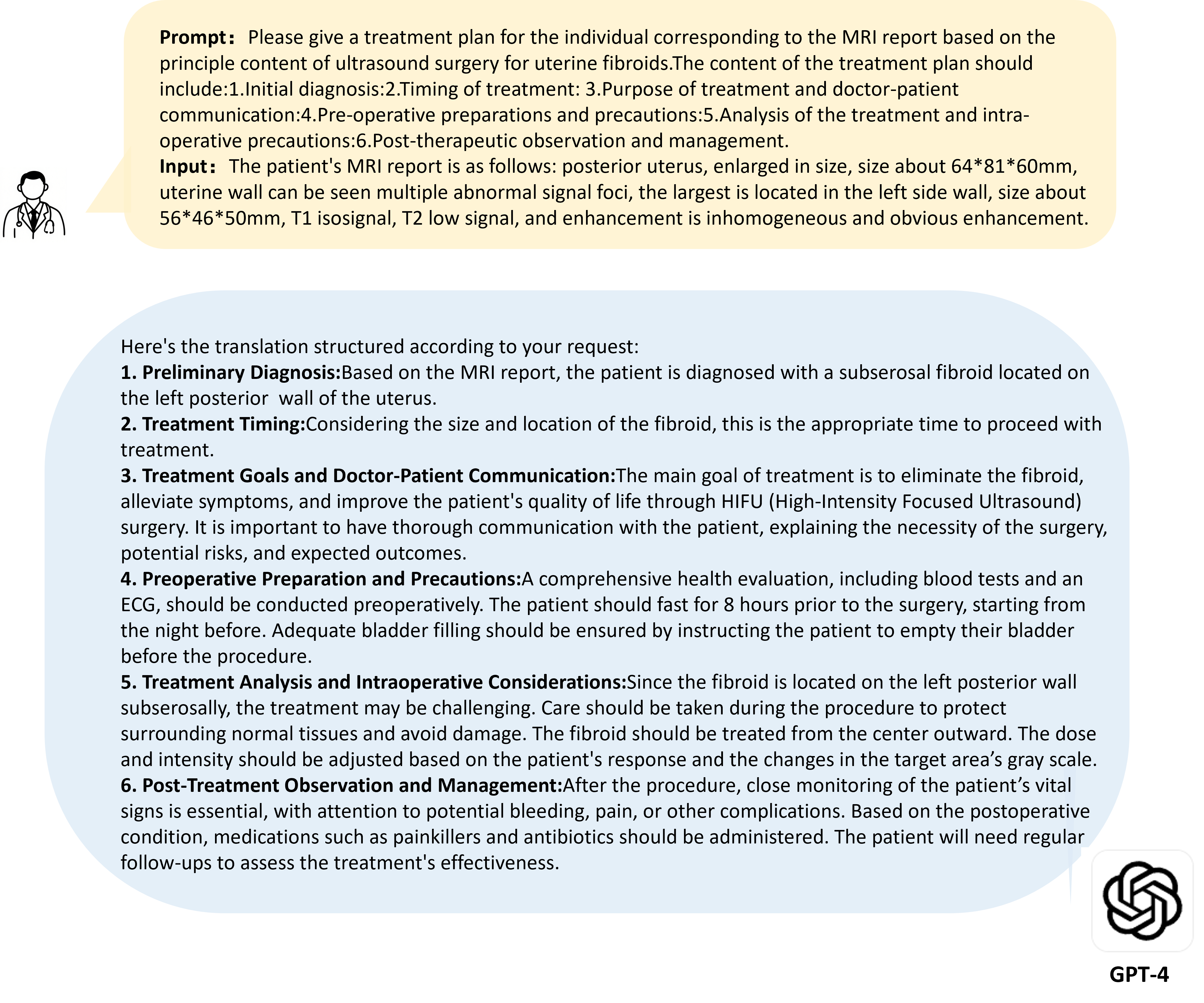}
    \caption{Treatment Plan Generated by GPT-4}
    \label{fig:appendixB_3}
\end{figure}
\clearpage

\begin{figure}[p]
    \centering
    \includegraphics[width=1.0\textwidth]{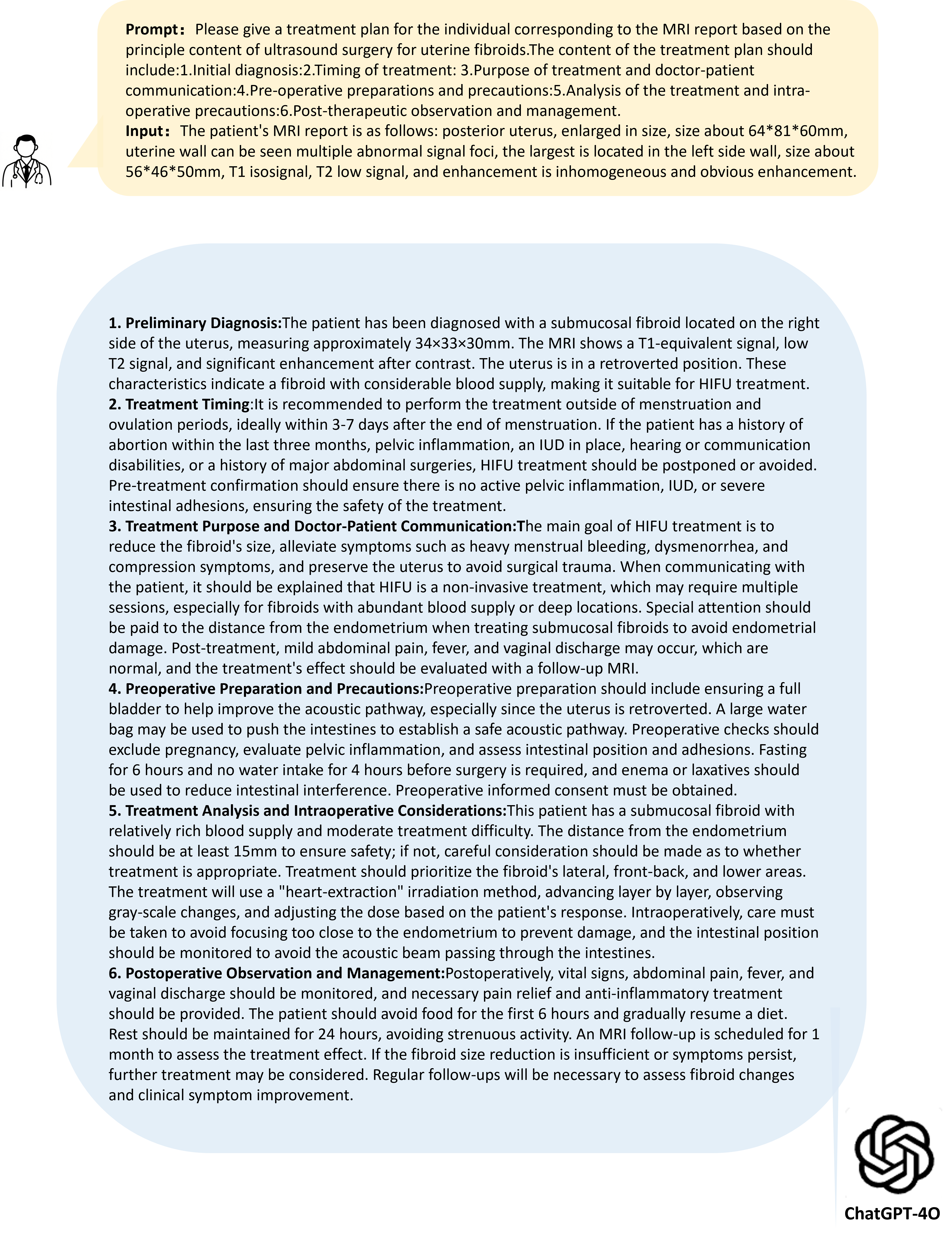}
    \caption{Treatment Plan Generated by ChatGPT-4o}
    \label{fig:appendixB_4}
\end{figure}
\clearpage

\begin{figure}[p]
    \centering
    \includegraphics[width=1.0\textwidth]{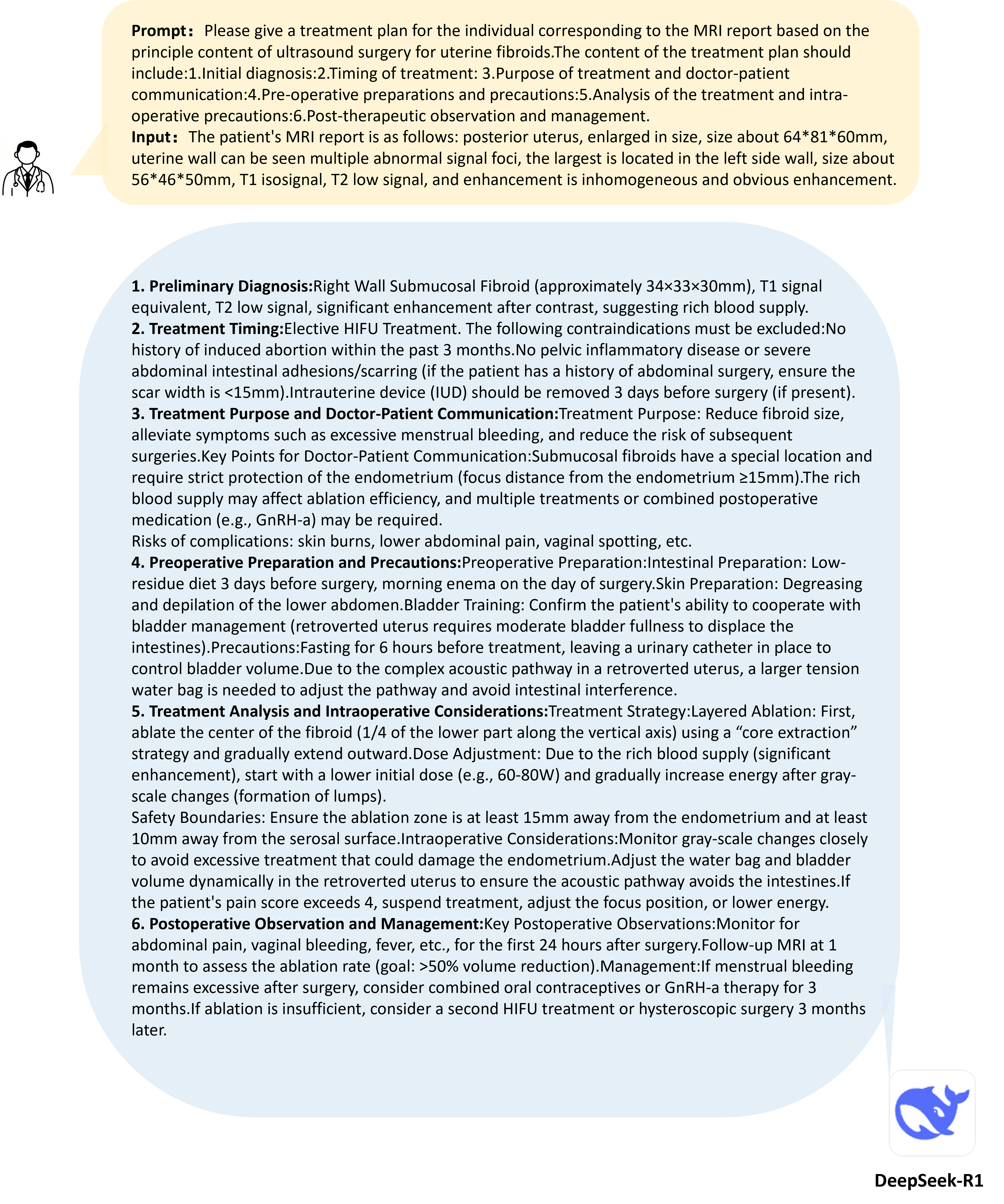}
    \caption{Treatment Plan Generated by DeepSeek}
    \label{fig:appendixB_5}
\end{figure}
\clearpage

\begin{figure}[p]
    \centering
    \includegraphics[width=1.0\textwidth, height=0.9\textheight, keepaspectratio]{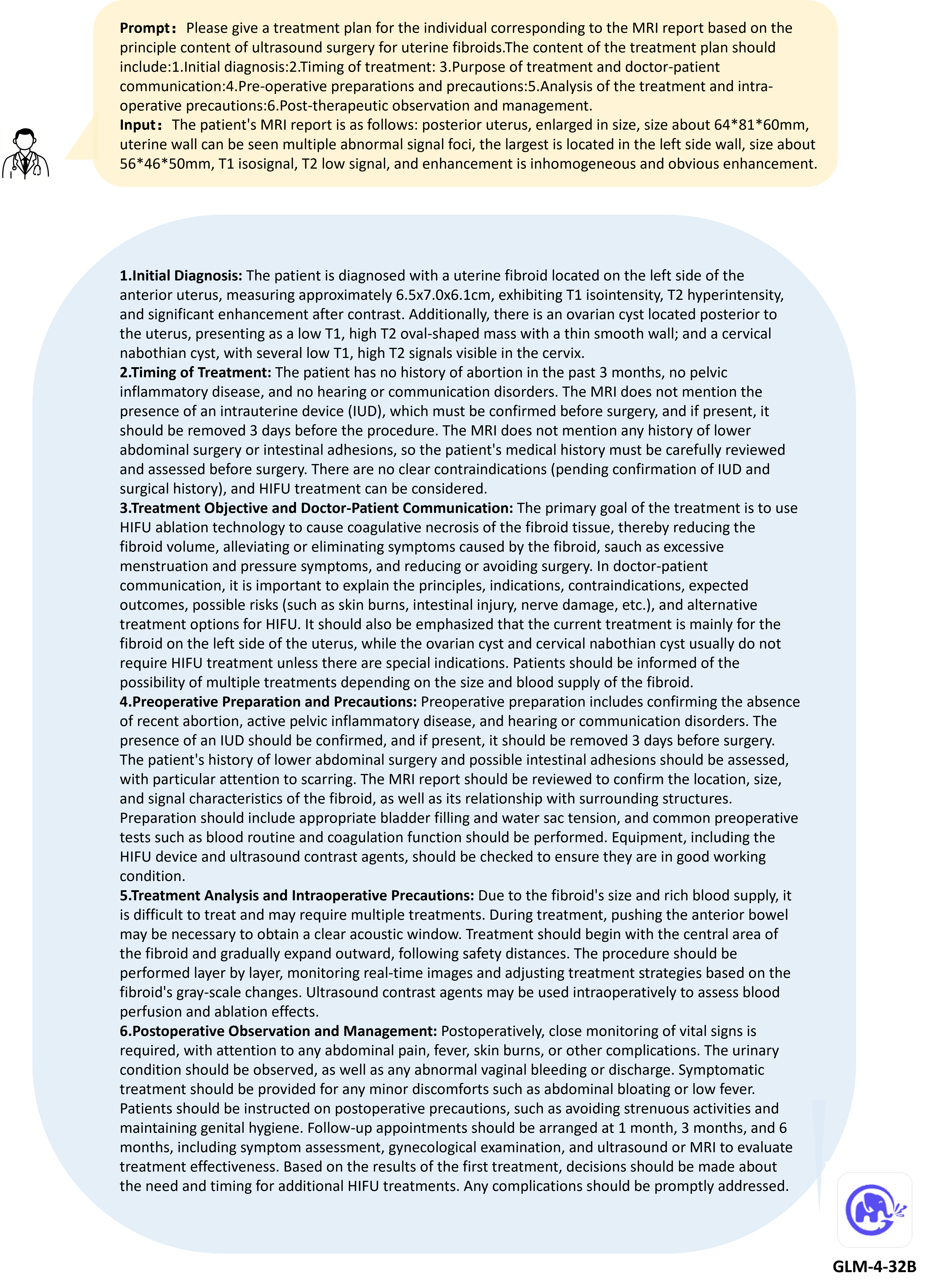}
    \caption{Treatment Plan Generated by GLM}
    \label{fig:appendixB_6}
\end{figure}
\clearpage

\section{Ablation Study Results}
\label{appendix:C}

\renewcommand{\thetable}{D\arabic{table}}
\setcounter{table}{0}

\begin{table}[htbp]
    \centering

    \label{tab:appendix_C}
    \begin{adjustbox}{max width=\textwidth}
    \begin{tabular}{lcccc}
\toprule
         Group & Completeness & Accuracy & Fluency & Compliance \\
\midrule
Fuul-Function Group (Baseline) & 82.5\%  & 80.0\% & 87.5\% & 97.5\% \\
Ablation Group 1 (No Executor) & 36.3\%  & 77.5\%    & 82.5\% & 92.5\% \\
Ablation Group 2 (No Optimizer)    & 82.5\% & 65.0\% & 77.5\%  & 72.5\% \\
Ablation Group 3 (No Memory)    & 76.3\% & 58.8\% & 73.8\%  & 65.0\% \\
\bottomrule
\end{tabular}
\end{adjustbox}
\end{table}

\section{Efficiency Analysis of FUAS-Agents System}
\label{appendix:D}

\renewcommand{\thetable}{D\arabic{table}}
\setcounter{table}{0}

\begin{table}[htbp]
    \centering

    \label{tab:appendix_D}
    \begin{adjustbox}{max width=\textwidth}
    \begin{tabular}{lcccc}
\toprule
          & Planner Agent & Executor Agent  & Strategy Agent & Optimizer Agent \\
\midrule
Running Times & 1.3  & 29.43 & 66.70 & 18.73 \\
Token Usage (k) & 131  & 0    & 6334 & 3178 \\
Success Rate    & 100\% & 100\% & 75\%  & 100\% \\
\bottomrule
\end{tabular}
\end{adjustbox}
\end{table}

\section{Valuation results from four senior FUAS clinicians}
\label{appendix:E}

\renewcommand{\thefigure}{E\arabic{figure}}
\setcounter{figure}{0}

\begin{figure}[ht]
    \centering
    \includegraphics[width=0.8\textwidth]{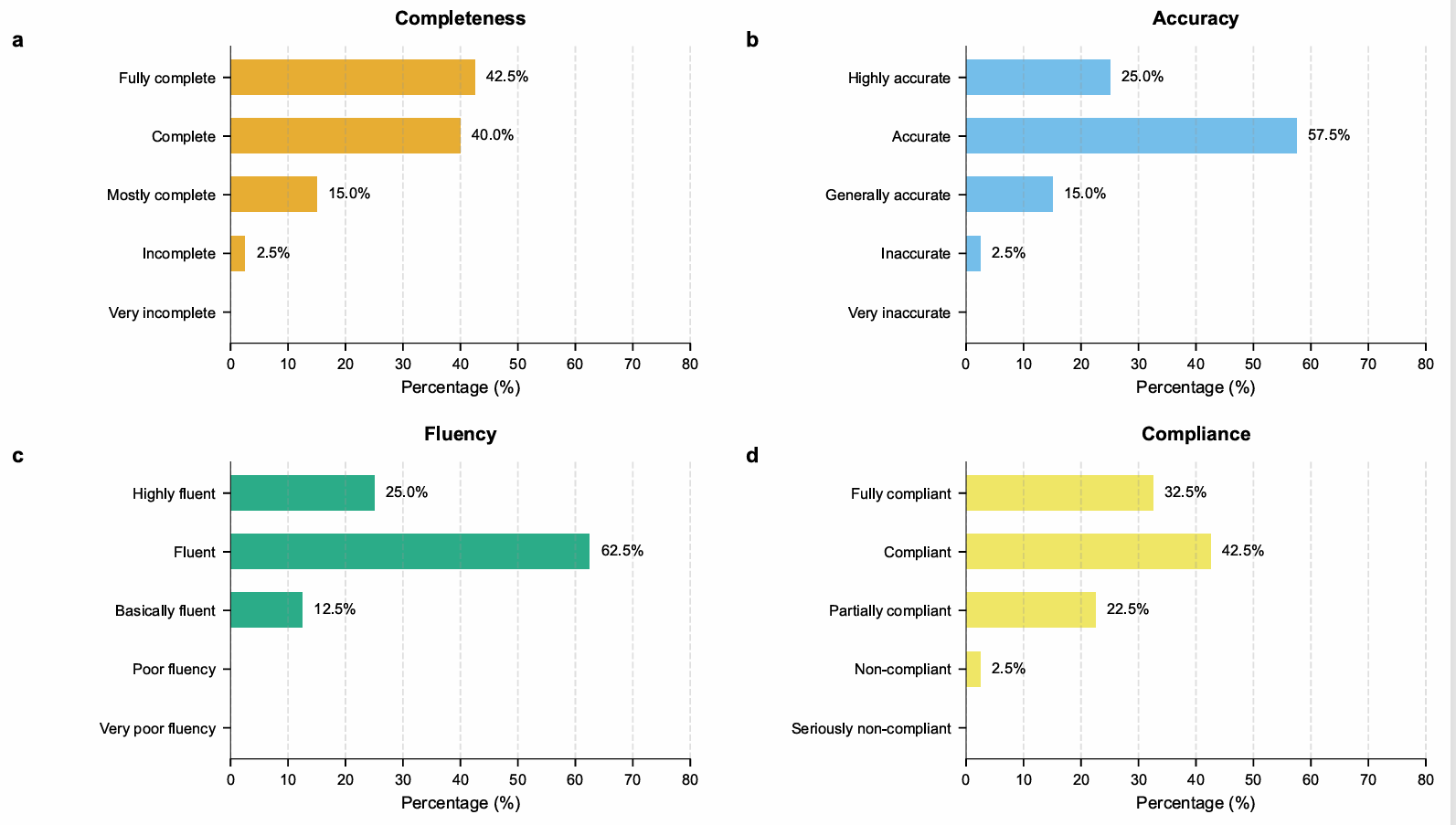}
    \caption{Valuation results from four senior FUAS clinicians}
    \label{fig:appendix_E}
\end{figure}

\clearpage
\section{Explanations of Formula Symbols}
\label{appendix:F}

\renewcommand{\thetable}{F\arabic{table}}
\setcounter{table}{0}

\begin{table}[ht]
  \centering
  \caption{Notation Definitions}
  \label{tab:appendix_F1}
  \begin{adjustbox}{max width=\textwidth}
  \begin{tabular}{p{0.25\textwidth}p{0.65\textwidth}}
    \toprule
    Notation & Description \\
    \midrule
    M2F, T2F, F2I & Transformation of different data types through the fine-tuning process. \\
    M2F & Maps MRI data (M) and patient metadata to features (F) for treatment planning. \\
    T2F & Maps tool outputs (e.g., segmentation results, dose predictions) to features (F). \\
    F2I & Transforms feature representations (F) into interpretable outputs (I), such as final treatment recommendations. \\
    \bottomrule
  \end{tabular}
  \end{adjustbox}
\end{table}

\begin{table}[ht]
  \centering
  \caption{Data Types and Structures}
  \label{tab:appendix_F2}
  \begin{adjustbox}{max width=\textwidth}
  \begin{tabular}{p{0.18\textwidth}p{0.37\textwidth}p{0.37\textwidth}}
    \toprule
    Data Type & Input Description & Output Description \\
    \midrule
    Medical Image Data (M) & 3D MRI image data with dimensions $(3 \times 25 \times 1024 \times 1024)$ & Processed image data generating segmentation masks or feature representations (F) for treatment planning. \\
    Dose Prediction Data & Input: predicted dose (float), weight (joblib), MRI image (nii), and mask (nii). & Output: Predicted dose values and associated weights. \\
    Clinical Data (B) & Structured patient data with $\sim$50 features (demographics, medical history). & Integrated data for the Strategy Agent. \\
    Feature Vector (F) & High-dimensional features (512 or 1024 dims) extracted from M or T. & Input for dose prediction or treatment planning. \\
    Treatment Plan (T) & Structured parameters (10 key parameters: dose, area, etc.). & Final treatment plan. \\
    Clinical Guidelines (G) & Expert-defined treatment standards and rules. & Ensures optimized, compliant treatment plans. \\
    Interpretative Output (I) & Model-generated treatment plans. & Final human-readable recommendations or reports. \\
    \bottomrule
  \end{tabular}
  \end{adjustbox}
\end{table}


\clearpage
\section{Model Performance Comparison}
\label{appendix:G}

\renewcommand{\thetable}{H\arabic{table}}
\setcounter{table}{0}

\begin{table}[ht]
  \centering
  \caption{Model Performance Comparison}
  \label{tab:appendix_G}
  \begin{tabular}{lll}
    \toprule
    Model Set & Model Name  & AUC \\
    \midrule
    Training    & Lasso-XGBoost Radiomics  & 0.97   \\
    & Logistic Regression Radiomics  & 0.73  \\
    & Support Vector Regressor Radiomics  & 0.85  \\
    & K-Nearest Neighbors Radiomics & 0.94  \\
    \midrule
    Test    & Lasso-XGBoost Radiomics  & 0.86   \\
    & Logistic Regression Radiomics  & 0.71   \\
    & Support Vector Regressor Radiomics  & 0.61   \\
    & K-Nearest Neighbors Radiomics  & 0.66   \\
    \bottomrule
  \end{tabular}
\end{table}